\documentclass[journal]{IEEEtran}

\usepackage{float}

\usepackage{cite}
\ifCLASSINFOpdf
  \usepackage[pdftex]{graphicx}
\else
\fi
\usepackage{color}
\usepackage{caption}
\usepackage{subcaption}

\usepackage[cmex10]{amsmath}
\usepackage{amsthm}
\usepackage{amssymb}

\usepackage{multirow}
\usepackage{flushend}

\usepackage{tikz}
\usetikzlibrary{shapes}

\usepackage{pgfplots}

\usepackage{stfloats}

\usepackage{threeparttable}

\newtheorem{proposition}{Proposition}

\newtheorem{property}{Property}
\theoremstyle{definition}
\newtheorem{definition}{Definition}

\newcommand{\Enf}[2]{E_{#1}^{(\mathbf{#2})}}
\newcommand{\E}[2]{E_{#1,\delta}^{(\mathbf{#2})}}
\newcommand{\Ex}[1]{\mathbb{E}\left[{#1}\right]}
\renewcommand{\Pr}[1]{\mathbb{P}\left[{#1}\right]}
\newcommand{\D}[2]{\Delta_{#1}^{(\mathbf{#2})}}
\newcommand{\C}[3]{C_{#1,\delta}^{(\mathbf{#2},\mathbf{#3})}}
\newcommand{\Cnf}[3]{C_{#1}^{(\mathbf{#2},\mathbf{#3})}}

\newcommand{\Znf}[2]{Z_{#1}^{(\mathbf{#2})}}
\newcommand{\Z}[2]{Z_{#1,\delta}^{(\mathbf{#2})}}

\newcommand{\W}[2]{W_{#1,\delta}^{(\mathbf{#2})}}

\definecolor{darkgreen}{rgb}{0,0.7,0}

\newcommand{\figurewidth}{0.95\columnwidth}
\newcommand{\figureheight}{0.78\columnwidth}

\newif\ifproofsend
\proofsendtrue

\begin{document}
\title{Faulty Successive Cancellation Decoding of Polar Codes for the Binary Erasure Channel}

\author{%
  Alexios~Balatsoukas-Stimming and~Andreas~Burg% 
  \thanks{A. Balatsoukas-Stimming and A. Burg are with the Telecommunications
    Circuits Laboratory (TCL), EPFL. Their research is supported by the Swiss
  National Science Foundation grant 200021\_149447.}% 
  \thanks{Part of this work has been presented at the International Symposium on 
		Information Theory and Its Applications (ISITA 2014)~\cite{Balatsoukas2014b}.}%
	\thanks{The authors would like to thank Mani Bastani Parizi for useful discussions.}%
}%
%\author{\authorblockN{Author names and affiliations omitted for blind review.}}
%\thanks{A. Balatsoukas-Stimming is with the Department of Electrical and Computer Engineering, Georgia Institute of Technology, Atlanta, GA, 30332 USA e-mail: (see http://www.michaelshell.org/contact.html).}% <-this % stops a space


\maketitle

\begin{abstract}
In this paper, faulty successive cancellation decoding of polar codes for the binary erasure channel is studied. To this end, a simple erasure-based fault model is introduced to represent errors in the decoder and it is shown that, under this model, polarization does not happen, meaning that fully reliable communication is not possible at any rate. Furthermore, a lower bound on the frame error rate of polar codes under faulty SC decoding is provided, which is then used, along with a well-known upper bound, in order to choose a blocklength that minimizes the erasure probability under faulty decoding. Finally, an unequal error protection scheme that can re-enable asymptotically erasure-free transmission at a small rate loss and by protecting only a constant fraction of the decoder is proposed. The same scheme is also shown to significantly improve the finite-length performance of the faulty successive cancellation decoder by protecting as little as $1.5$\% of the decoder.
\end{abstract}

\begin{IEEEkeywords}
Polar codes, successive cancellation decoding, faulty decoding.
\end{IEEEkeywords}

% \fi
\IEEEpeerreviewmaketitle

\section{Introduction}\label{sec:introduction}
%TODO:
%\begin{enumerate}
	%\item Can we show something about the correlation coefficients in this case?
	%\item Use bounds to get optimal blocklength. If LB of one length overlaps with UB of other length, we can not say much. In this case, we can use the bounds to narrow down the candidate codes, and then simulate to get the best one.
	%\item Do simulations to verify bounds (mainly illustration, nothing that is strictly required).
%\end{enumerate}

\IEEEPARstart{U}{ncertainties} in the manufacturing process of integrated circuits are expected to play a significant role in the design of very-large-scale integration systems in the nanoscale era \cite{Borkar2005,Unsal2006,Ghosh2010}. Due to these uncertainties, it will become more and more difficult to guarantee the correct behavior of integrated circuits at the gate level, meaning that the hardware may become \emph{faulty} in the sense that data is not always processed or stored correctly. Moreover, very aggressive dynamic voltage scaling, which is commonly used to reduce the energy consumption of integrated circuits, can increase the occurrence of undesired faulty behavior~\cite{Bhavnagarwala2005}. Traditional methods to ensure accurate hardware behavior, such as using larger transistors or circuit-level error correcting codes, are costly in terms of both area and power. 

Fortunately, many applications are inherently fault tolerant in the sense that they do not fail catastrophically under faulty hardware. A good example of such an application are communication systems, and more specifically channel decoders, since the processed data is already probabilistic in nature due to transmission over a noisy channel. Faulty iterative decoding of LDPC codes was first studied in~\cite{Varshney2011}, where the Gallager A and sum-product algorithms are considered. Later studies also targeted the Gallager B algorithm~\cite{Yazdi2013,Leduc-Primeau2012,Sala2017}, the min-sum algorithm~\cite{Kameni2013,Balatsoukas2014,KameniNgassa2015}, as well as more general message-passing algorithms~\cite{Huang2015,Dupraz2015}. All of the aforementioned studies provide valuable insight into the limitations of LDPC codes under various decoding algorithms and fault models. Unfortunately, in many cases, the conclusion is that fully reliable communication is not possible when faults are present inside the decoder itself. Surprisingly, in some special cases, it has been demonstrated that faulty decoders can in fact even improve the error performance of LDPC codes in the finite blocklength regime~\cite{Ngassa2014,AlRasheed2014,Vasic2015,Ivanis2016}. LDPC codes are usually studied with the help of random ensembles, meaning that a family of codes is studied rather than individual codes. Moreover there exists an infinite number of code ensembles with a given coding rate. Thus, it becomes unclear which code ensemble and which individual code should be studied.

Polar codes~\cite{Arikan2009} constitute a different class of channel codes which has recently attracted significant attention, albeit not yet in the context of faulty decoding. Contrary to LDPC codes, a polar code for a given channel and coding rate is uniquely defined, thus greatly simplifying the choice of code to examine. Polar codes are provably capacity achieving over various channels and they have an efficient and structured successive cancellation (SC) decoding algorithm whose complexity is $O(N\log N)$, where $N$ is the length of the code. Moreover, encoding can also be performed with complexity that is $O(N\log N)$. When used for transmission over the binary erasure channel (BEC), the SC decoding algorithm can be highly simplified. Moreover, there exist analytical upper and lower bounds on the frame erasure rate (FER), which have been shown to be tight~\cite{Bastani2013} and enable us to have a very good approximation of the FER without resorting to lengthy Monte-Carlo simulations.

\subsubsection*{Contribution}
In this paper we study SC decoding of polar codes for transmission over the BEC under an erasure-based internal fault model. We show that, under the fault model assumed in this paper, fully reliable communication is no longer possible. Furthermore, by studying the polarization process, we show that synthetic channel ordering with respect to both the channel erasure probability and the internal decoder erasure probability holds. We also adapt the lower bound on the FER derived in~\cite{Bastani2013} to the case of faulty decoding, and we use it in order to derive the FER-optimal blocklength for a polar code of a given rate, and for a given channel and decoder erasure probability. Finally, we introduce a simple unequal error protection method, which is shown to re-enable asymptotically fully reliable communication by protecting only a constant fraction of the decoder. In the finite blocklength regime, our proposed fault-tolerance method significantly improves the FER with very low overhead. 

\subsubsection*{Outline}
The remainder of this paper is organized as follows. Section~\ref{sec:background} provides some background on the construction and decoding of polar codes. In Section~\ref{sec:faulty}, we introduce the fault model that is used throughout this paper and we prove that fully reliable communication using polar codes is not possible under faulty decoding over the BEC. We also show some other useful properties of the faulty decoder. Moreover, in Section~\ref{sec:fer} we adapt the lower bound on the FER derived in~\cite{Bastani2013} to the case of faulty decoding, and in Section~\ref{sec:protection} we describe our proposed unequal error protection scheme. A discussion and some results on the optimal blocklength under faulty decoding are provided in Section~\ref{sec:optbl}. In Section~\ref{sec:numerical}, we show numerical results concerning the FER, the optimal choice of blocklength, as well as the effectiveness of our proposed unequal error protection method. Finally, Section~\ref{sec:conclusion} concludes this paper.

\subsubsection*{Notation} 
We use the notation $\overline{X} \triangleq 1 - X$. We use boldface letters to denote vectors, matrices, and strings. The $n$-th character of a string $\mathbf{s}$ is denoted by $\mathbf{s}_n$. We use $\log(\cdot)$ to denote the binary logarithm. We denote the binary erasure channel with erasure probability $p$ as BEC$(p)$ and the ternary erasure channel with erasure probability $p$ as TEC$(p)$. We use $\emptyset$ to denote an empty string. Finally, we use $|\cdot|$ to denote both the length of a string and the cardinality of a set. We use $\lfloor x \rceil$ to denote the nearest integer of $x$ (i.e., the rounding function) and $\lceil x \rceil$ to denote the ceiling function.

\section{Polar Codes}\label{sec:background}

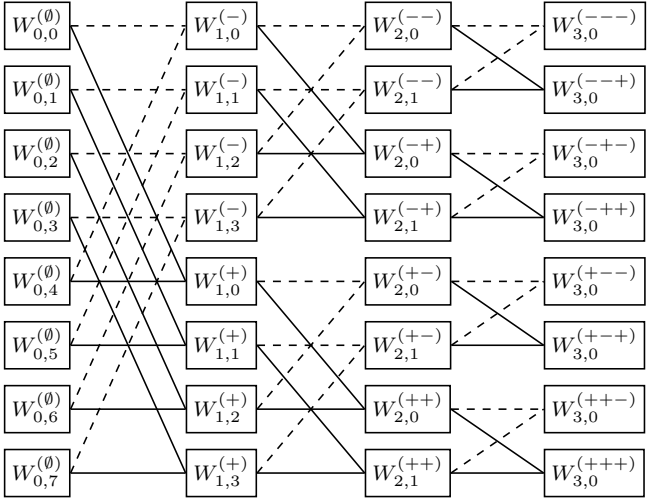
\begin{figure}[t]
\centering
\begin{tikzpicture}[scale=0.85]
	\footnotesize
	\node[rectangle, draw, semithick, left] at (0,0) (l00) {$W^{(\emptyset)}_{0,7}$};
	\node[rectangle, draw, semithick, left] at (0,1) (l01) {$W^{(\emptyset)}_{0,6}$};
	\node[rectangle, draw, semithick, left] at (0,2) (l02) {$W^{(\emptyset)}_{0,5}$};
	\node[rectangle, draw, semithick, left] at (0,3) (l03) {$W^{(\emptyset)}_{0,4}$};
	\node[rectangle, draw, semithick, left] at (0,4) (l04) {$W^{(\emptyset)}_{0,3}$};
	\node[rectangle, draw, semithick, left] at (0,5) (l05) {$W^{(\emptyset)}_{0,2}$};
	\node[rectangle, draw, semithick, left] at (0,6) (l06) {$W^{(\emptyset)}_{0,1}$};
	\node[rectangle, draw, semithick, left] at (0,7) (l07) {$W^{(\emptyset)}_{0,0}$};

	\node[rectangle, draw, semithick, right] at (1.8,0) (l10) {$W^{(+)}_{1,3}$};
	\node[rectangle, draw, semithick, right] at (1.8,1) (l11) {$W^{(+)}_{1,2}$};
	\node[rectangle, draw, semithick, right] at (1.8,2) (l12) {$W^{(+)}_{1,1}$};
	\node[rectangle, draw, semithick, right] at (1.8,3) (l13) {$W^{(+)}_{1,0}$};
	\node[rectangle, draw, semithick, right] at (1.8,4) (l14) {$W^{(-)}_{1,3}$};
	\node[rectangle, draw, semithick, right] at (1.8,5) (l15) {$W^{(-)}_{1,2}$};
	\node[rectangle, draw, semithick, right] at (1.8,6) (l16) {$W^{(-)}_{1,1}$};
	\node[rectangle, draw, semithick, right] at (1.8,7) (l17) {$W^{(-)}_{1,0}$};

	\node[rectangle, draw, semithick, right] at (4.6,0) (l20) {$W^{(++)}_{2,1}$};
	\node[rectangle, draw, semithick, right] at (4.6,1) (l21) {$W^{(++)}_{2,0}$};
	\node[rectangle, draw, semithick, right] at (4.6,2) (l22) {$W^{(+-)}_{2,1}$};
	\node[rectangle, draw, semithick, right] at (4.6,3) (l23) {$W^{(+-)}_{2,0}$};
	\node[rectangle, draw, semithick, right] at (4.6,4) (l24) {$W^{(-+)}_{2,1}$};
	\node[rectangle, draw, semithick, right] at (4.6,5) (l25) {$W^{(-+)}_{2,0}$};
	\node[rectangle, draw, semithick, right] at (4.6,6) (l26) {$W^{(--)}_{2,1}$};
	\node[rectangle, draw, semithick, right] at (4.6,7) (l27) {$W^{(--)}_{2,0}$};

	\node[rectangle, draw, semithick, right] at (7.4,0) (l30) {$W^{(+++)}_{3,0}$};
	\node[rectangle, draw, semithick, right] at (7.4,1) (l31) {$W^{(++-)}_{3,0}$};
	\node[rectangle, draw, semithick, right] at (7.4,2) (l32) {$W^{(+-+)}_{3,0}$};
	\node[rectangle, draw, semithick, right] at (7.4,3) (l33) {$W^{(+--)}_{3,0}$};
	\node[rectangle, draw, semithick, right] at (7.4,4) (l34) {$W^{(-++)}_{3,0}$};
	\node[rectangle, draw, semithick, right] at (7.4,5) (l35) {$W^{(-+-)}_{3,0}$};
	\node[rectangle, draw, semithick, right] at (7.4,6) (l36) {$W^{(--+)}_{3,0}$};
	\node[rectangle, draw, semithick, right] at (7.4,7) (l37) {$W^{(---)}_{3,0}$};

	\draw[semithick] (l00.east) -- (l10.west);
	\draw[semithick] (l01.east) -- (l11.west);
	\draw[semithick] (l02.east) -- (l12.west);
	\draw[semithick] (l03.east) -- (l13.west);
	\draw[semithick,dashed] (l04.east) -- (l14.west);
	\draw[semithick,dashed] (l05.east) -- (l15.west);
	\draw[semithick,dashed] (l06.east) -- (l16.west);
	\draw[semithick,dashed] (l07.east) -- (l17.west);
	\draw[semithick,dashed] (l00.east) -- (l14.west);
	\draw[semithick,dashed] (l01.east) -- (l15.west);
	\draw[semithick,dashed] (l02.east) -- (l16.west);
	\draw[semithick,dashed] (l03.east) -- (l17.west);
	\draw[semithick] (l04.east) -- (l10.west);
	\draw[semithick] (l05.east) -- (l11.west);
	\draw[semithick] (l06.east) -- (l12.west);
	\draw[semithick] (l07.east) -- (l13.west);

	\draw[semithick] (l10.east) -- (l20.west);
	\draw[semithick] (l11.east) -- (l21.west);
	\draw[semithick,dashed] (l12.east) -- (l22.west);
	\draw[semithick,dashed] (l13.east) -- (l23.west);
	\draw[semithick] (l14.east) -- (l24.west);
	\draw[semithick] (l15.east) -- (l25.west);
	\draw[semithick,dashed] (l16.east) -- (l26.west);
	\draw[semithick,dashed] (l17.east) -- (l27.west);
	\draw[semithick,dashed] (l10.east) -- (l22.west);
	\draw[semithick,dashed] (l11.east) -- (l23.west);
	\draw[semithick] (l12.east) -- (l20.west);
	\draw[semithick] (l13.east) -- (l21.west);
	\draw[semithick,dashed] (l14.east) -- (l26.west);
	\draw[semithick,dashed] (l15.east) -- (l27.west);
	\draw[semithick] (l16.east) -- (l24.west);
	\draw[semithick] (l17.east) -- (l25.west);

	\draw[semithick] (l20.east) -- (l30.west);
	\draw[semithick,dashed] (l21.east) -- (l31.west);
	\draw[semithick] (l22.east) -- (l32.west);
	\draw[semithick,dashed] (l23.east) -- (l33.west);
	\draw[semithick] (l24.east) -- (l34.west);
	\draw[semithick,dashed] (l25.east) -- (l35.west);
	\draw[semithick] (l26.east) -- (l36.west);
	\draw[semithick,dashed] (l27.east) -- (l37.west);
	\draw[semithick,dashed] (l20.east) -- (l31.west);
	\draw[semithick] (l21.east) -- (l30.west);
	\draw[semithick,dashed] (l22.east) -- (l33.west);
	\draw[semithick] (l23.east) -- (l32.west);
	\draw[semithick,dashed] (l24.east) -- (l35.west);
	\draw[semithick] (l25.east) -- (l34.west);
	\draw[semithick,dashed] (l26.east) -- (l37.west);
	\draw[semithick] (l27.east) -- (l36.west);

\end{tikzpicture}
\caption{Synthetic channel construction for a polar code of length $N = 2^3 = 8$. Pairs of solid lines represent the $+$ transformation and pairs of dashed lines represent the $-$ transformation.}\label{fig:polartransform}
\end{figure}

\subsection{Polarizing Channel Transformation}
Let $W$ denote a binary input memoryless channel with input $u~\in~\{0,1\}$, output $y~\in~\mathcal{Y}$, and transition probabilities $W(y|u)$. The polarizing transformation proposed by Ar{\i}kan~\cite[Section I]{Arikan2009} generates $N \triangleq 2^n$ \emph{synthetic channels} in $n$ steps as follows. At step $1$ of the polarizing transformation, $N$ independent copies of the channel $W$, denoted by $W^{(\emptyset)}_{0,k},~k = 0,\hdots,N-1,$ are combined pair-wise in order to generate $N/2$ independent copies of a pair of new synthetic channels denoted by $W_{1,k}^{(+)}$ and $W_{1,k}^{(-)},~k = 0,\hdots,N/2-1$. The ``+'' channels can be shown to be better, in terms of mutual information and Bhattacharyya parameter, than the original channel, while the ``-'' channels are worse than the original channel. The same transformation is applied to $W_{1,k}^{(+)}$ and $W_{1,k}^{(-)},~k = 0,\hdots,N/2-1$ in order to generate $N/4$ independent copies of $W_{2,k}^{(++)}$, $W_{2,k}^{(+-)}$, $W_{2,k}^{(-+)}$ and $W_{2,k}^{(--)},~k = 0,\hdots,N/4-1$. This procedure is repeated for a total of $n$ steps, until $2^n$ channels $W_{n,0}^{(\mathbf{s})},~\mathbf{s}\in\{+,-\}^n,$ are generated. Note that, in general, the notation $W_{s,k}^{(\mathbf{s})}$ implies that $|\mathbf{s}| = s$ and for this reason we have $\mathbf{s}\in\{+,-\}^n$ for the final combining step where $s=n$. An example of the transformation steps is depicted in Figure~\ref{fig:polartransform} for $n = 3$.

\subsection{Erasure Probability of Synthetic Channels}
Let $Z_{s,k}^{(\mathbf{s})} \triangleq Z\left(W_{s,k}^{(\mathbf{s})}\right)$ denote the Bhattacharyya parameter of the synthetic channel $W_{s,k}^{(\mathbf{s})}$. When $W$ is a BEC$(p)$, its Bhattacharyya parameter is equal to the erasure probability, i.e., $Z\left(W_{0,k}^{(\emptyset)}\right) = Z(W) = p$. Moreover, all synthetic channels generated at step $s$ are also BECs and their Bhattacharyya parameters (equivalently, their erasure probabilities) can be calculated recursively based on the Bhattacharyya parameters of the channels at step $(s-1)$ as~\cite[Section III]{Arikan2009}
\begin{align}
	Z_{s,k}^{(\mathbf{s-})}	& = Z_{s-1,k}^{(\mathbf{s})} + Z_{s-1,k+2^{n-s}}^{(\mathbf{s})} - Z_{s-1,k}^{(\mathbf{s})}Z_{s-1,k+2^{n-s}}^{(\mathbf{s})}, \label{eq:zup1full} \\
	Z_{s,k}^{(\mathbf{s+})}	& = Z_{s-1,k}^{(\mathbf{s})}Z_{s-1,k+2^{n-s}}^{(\mathbf{s})}, \label{eq:zup2full} 
\end{align}
where $s = 1,\hdots,n,~k = 0,\hdots,2^{n-s}-1$. The channels $W^{(\mathbf{s})}_{s,k},~k = 0, \hdots, 2^{n-s}-1,$ are independent copies of the same type of channel, meaning that their erasure probabilities are identical. Thus, if we are only interested in the erasure probability of a specific \emph{type} $\mathbf{s}$ of channel we can simplify~\eqref{eq:zup1full} and~\eqref{eq:zup2full} by omitting the index $k$ as
\begin{align}
	Z_{s}^{(\mathbf{s-})}	& = T^{-}\left(Z_{s-1}^{(\mathbf{s})}\right) \triangleq 2Z_{s-1}^{(\mathbf{s})} - \left(Z_{s-1}^{(\mathbf{s})}\right)^2, \label{eq:zup1} \\
	Z_{s}^{(\mathbf{s+})}	& = T^{+}\left(Z_{s-1}^{(\mathbf{s})}\right) \triangleq \left(Z_{s-1}^{(\mathbf{s})}\right)^2, \label{eq:zup2} 
\end{align}
with $Z_{0}^{(\emptyset)} = p$. The vector containing all $Z^{(\mathbf{s})}_{s},~\mathbf{s}\in\{+,-\}^s,$ variables is denoted by $\mathbf{Z}_{s}$.

Moreover, as in \cite{Arikan2009,Hassani2012}, we define the polarization random process $\epsilon _s$ as
\begin{align}
	\epsilon _{s} = \Znf{s}{s},
\end{align}with $\Pr{\mathbf{S} = \mathbf{s}} = \frac{1}{2^s}$, i.e., $\epsilon _{s}$ is equally likely to be equal to the erasure probability of any of the $2^s$ distinct types of synthetic channels at step $s$ of the polarizing transformation. The random process $\epsilon _s$ can be written equivalently as
\begin{align}
	\epsilon _{s} = \left\{ \begin{array}{ll} T^{-}(\epsilon_{s-1}) & \text{w.p. }~1/2, \\ T^{+}(\epsilon_{s-1}) & \text{w.p. }~1/2,  \end{array} \right. \label{eq:nonfaulty}
\end{align}
with $\epsilon _0 = Z(W) = p$. It was shown in \cite[Theorem 1]{Arikan2009} that $\epsilon _s$ converges almost surely to a random variable $\epsilon _{\infty} \in \{0,1\}$, with $P(\epsilon _{\infty} = 0) = I(W) = 1 -p$, where $I(W)$ denotes the symmetric capacity of the BEC $W$.

Finally, let us define a binary erasure indicator variable $E_{s,k}^{(\mathbf{s})}$ for which $E_{s,k}^{(\mathbf{s})} = 1$ if and only if the output of the synthetic channel $W_{s,k}^{(\mathbf{s})}$ is an erasure and $E_{s,k}^{(\mathbf{s})} = 0$ otherwise. It is clear that $\Ex{E_{s,k}^{(\mathbf{s})}} = Z_{s,k}^{(\mathbf{s})}$. The indicator variables can also be determined recursively as follows~\cite{Bastani2013}
\begin{align}
	E_{s,k}^{(\mathbf{s-})}	& = E_{s-1,k}^{(\mathbf{s})} + E_{s-1,k+2^{n-s}}^{(\mathbf{s})} - E_{s-1,k}^{(\mathbf{s})}E_{s-1,k+2^{n-s}}^{(\mathbf{s})}, \label{eq:indup1full}\\
	E_{s,k}^{(\mathbf{s+})}	& = E_{s-1,k}^{(\mathbf{s})}E_{s-1,k+2^{n-s}}^{(\mathbf{s})}. \label{eq:indup2full}
\end{align}
Similarly to the Bhattacharyya parameters, if we are only interested in the statistics of the indicator variable for a channel of a specific type $\mathbf{s}$, we can simplify~\eqref{eq:indup1full} and~\eqref{eq:indup2full} as
\begin{align}
	E_{s}^{(\mathbf{s-})}	& = {E_{s-1}^{(\mathbf{s})}}' + {E_{s-1}^{(\mathbf{s})}}'' - {E_{s-1}^{(\mathbf{s})}}'{E_{s-1}^{(\mathbf{s})}}'', \label{eq:indup1}\\
	E_{s}^{(\mathbf{s+})}	& = {E_{s-1}^{(\mathbf{s})}}'{E_{s-1}^{(\mathbf{s})}}'', \label{eq:indup2}
\end{align}
where ${E_{s-1}^{(\mathbf{s})}}'$ and ${E_{s-1}^{(\mathbf{s})}}''$ denote two independent realizations of $E_{s-1}^{(\mathbf{s})}$~\cite{Bastani2013}. The vector containing all $E^{(\mathbf{s})}_{s}$ indicator variables is denoted by $\mathbf{E}_{s}$.

\subsection{Construction of Polar Codes}
Let us define a mapping from $\textbf{s} \in \{+,-\}^n$ to the integer-valued indices $i \in \{0,\hdots,2^n-1\}$ as follows. First, we construct $\mathbf{b}$ by replacing each $-$ that appears in $\mathbf{s}$ with a $0$ and each $+$ that appears in $\mathbf{s}$ with a $1$. Then, the index $i$ can be obtained by considering $\textbf{b}$ as a left-MSB binary representation of $i$. As this mapping is a bijection, we use $\textbf{s}$ and $i$ interchangeably. 

Let us fix a blocklength $N = 2^n$ and a code rate $R \triangleq \frac{K}{N},~0 < K < N$. Moreover, let $\mathcal{A}$ denote the set of the $K$ channel indices $i$ (equivalently, strings $\mathbf{s}$) with the smallest $\Znf{n}{s}$. A polar code of rate $R$ is constructed by transmitting the information vector $\mathbf{u}_{\mathcal{A}}$ over the $K$ best synthetic channels, while freezing the input of the remaining synthetic channels, i.e., $\mathbf{u}_{\mathcal{A}^c}$ to a value that is known at the receiver. This can be achieved by transmitting the encoded codeword $\mathbf{x} = \mathbf{u}\mathbf{G}_n$ over $2^n$ independent uses of the initial BEC $W$, where
\begin{align}
	\mathbf{G}_n	& = \mathbf{B}_n\mathbf{F}^{\otimes n}, \qquad \mathbf{F} = \begin{bmatrix} 1 & 0 \\ 1 & 1 \end{bmatrix},
\end{align}
and $\mathbf{B}_n$ denotes the bit-reversal permutation matrix~\cite[Section VII-B]{Arikan2009}. Due to the structure of $\mathbf{G}_n$, encoding can be performed with complexity $O(N \log N)$. If $R < I(W) = 1 - p$, then as $n$ is increased, all synthetic channels $W_{n,0}^{(\mathbf{s})},~\mathbf{s} \in\mathcal{A},$ become arbitrarily good and the polar code is capacity achieving~\cite[Theorem 2]{Arikan2009}.

\subsection{Successive Cancellation Decoding of Polar Codes}
Without loss of generality, we assume the output alphabet of the BEC $W$ to be $\mathcal{Y} = \{-1,0,+1\}$, where $0$ denotes an erasure, while $-1$ corresponds to the binary input $1$ and $+1$ corresponds to the binary input $0$. The SC decoder proposed by Ar{\i}kan~\cite[Section VIII]{Arikan2009} decodes the synthetic channels $W_{n}^{(\mathbf{s})},~\mathbf{s} \in \mathcal{A}_n,$ successively following a natural ordering with respect to $i$ (this is equivalent to a top-down decoding order of the $W_{3,0}^{(\mathbf{s})},~s \in \{+,-\}^3$, channels in Figure~\ref{fig:polartransform}). The input of the channels $\mathbf{s} \notin \mathcal{A}_n$ does not need to be decoded as, by construction, it is known a-priori to the receiver. 

In order to estimate the input of the synthetic channel $W_{n}^{(\mathbf{s})}$, the $N$ channel outputs resulting from $N$ independent uses of $W$, i.e., the outputs of $W^{(\emptyset)}_{0,k},~k = 0,\hdots,N$, are combined pair-wise through a full binary tree structure of depth $n$ that is identical to the channel combining structure of Figure~\ref{fig:polartransform}. For each combining step, one of two possible update rules is used depending on the synthetic channel $\mathbf{s}$ and the stage $s$ that is being processed. More specifically, the two possible update rules are
\begin{align}
	f^-(m_1,m_2)	& = m_1m_2, \\
	f^+(m_1,m_2,u)	& = \left\lfloor\frac{(-1)^um_1 + m_2}{2}\right\rceil,
\end{align}
where $m_1,m_2 \in \{{-1},0,{+1}\}$ and $u$ denotes a \emph{partial sum}, which is the modulo-$2$ sum of some of the previously decoded bits.\footnote{We note that we use $\lfloor -0.5 \rceil = -1$ and $\lfloor 0.5 \rceil = 1$ for tie-breaking in $f^+$.} If $\mathbf{s}_s = -,$ then all updates at level $s$ of the tree are performed using $f^-$, while if $\mathbf{s}_s = +,$ then all updates at level $s$ of the tree are performed using $f^+$. The partial sums required by each of the $f^+$ nodes at level $s$ can be calculated from the partial sums at level $s+1$,~either recursively \cite[Proposition 3]{Arikan2009} or directly~\cite[Section VI-F]{Leroux2013}.  When level $n$ is reached, the output message will either be correct (i.e., ${-1}$ or ${+1}$), or an erasure. If the final output message is correct, we can derive the corresponding bit value for $\mathbf{u}_i$ and proceed with decoding. If the final output message is an erasure, the decoder halts and declares a block erasure. By re-using intermediate synthetic channel outputs, it can be shown that the complexity of SC decoding is $O(N \log N)$ \cite[Section VIII]{Arikan2009}.

\section{Faulty SC Decoding of Polar Codes}\label{sec:faulty}

All current SC decoder hardware implementations (e.g., \cite{Leroux2011,Mishra2012,Zhang2013,Leroux2013}) require a full binary tree of \emph{memory elements} (MEs) of depth $n$, which store the messages that result from the update rules at each level of the decoder tree. The total number of MEs required by a decoder is
\begin{align}
	N_{\text{ME}}	& = \sum _{s=0}^n2^{n-s} = 2^{n+1} -1 = 2N-1 \in O(N).
\end{align}
The processing elements (PEs), which apply the update rules, can also have a full binary tree structure for a fully-parallel implementation~\cite{Leroux2011}, although semi-parallel implementations are also possible~\cite{Leroux2013}. A fully-parallel implementation requires $N-1$ PEs, while in a semi-parallel implementation the number of PEs is restricted to $P < N - 1$. 

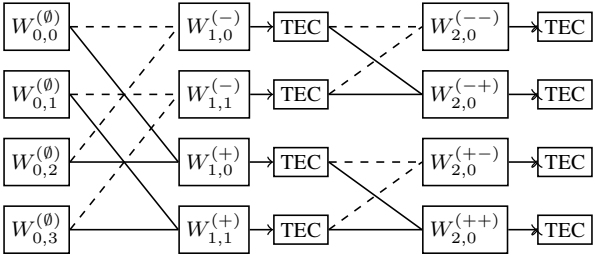
\begin{figure}[t]
\centering
\begin{tikzpicture}[scale=0.9]
	\footnotesize
	\node[rectangle, draw, semithick, left] at (0,0) (l00) {$W^{(\emptyset)}_{0,3}$};
	\node[rectangle, draw, semithick, left] at (0,1) (l01) {$W^{(\emptyset)}_{0,2}$};
	\node[rectangle, draw, semithick, left] at (0,2) (l02) {$W^{(\emptyset)}_{0,1}$};
	\node[rectangle, draw, semithick, left] at (0,3) (l03) {$W^{(\emptyset)}_{0,0}$};

	\node[rectangle, draw, semithick, right] at (1.6,0) (l10) {$W^{(+)}_{1,1}$};
	\node[rectangle, draw, semithick, right] at (1.6,1) (l11) {$W^{(+)}_{1,0}$};
	\node[rectangle, draw, semithick, right] at (1.6,2) (l12) {$W^{(-)}_{1,1}$};
	\node[rectangle, draw, semithick, right] at (1.6,3) (l13) {$W^{(-)}_{1,0}$};

	\node[rectangle, draw, semithick, right] at (3,0) (b10) {TEC};
	\node[rectangle, draw, semithick, right] at (3,1) (b11) {TEC};
	\node[rectangle, draw, semithick, right] at (3,2) (b12) {TEC};
	\node[rectangle, draw, semithick, right] at (3,3) (b13) {TEC};

	\node[rectangle, draw, semithick, right] at (5.2,0) (l20) {$W^{(++)}_{2,0}$};
	\node[rectangle, draw, semithick, right] at (5.2,1) (l21) {$W^{(+-)}_{2,0}$};
	\node[rectangle, draw, semithick, right] at (5.2,2) (l22) {$W^{(-+)}_{2,0}$};
	\node[rectangle, draw, semithick, right] at (5.2,3) (l23) {$W^{(--)}_{2,0}$};

	\node[rectangle, draw, semithick, right] at (6.9,0) (b20) {TEC};
	\node[rectangle, draw, semithick, right] at (6.9,1) (b21) {TEC};
	\node[rectangle, draw, semithick, right] at (6.9,2) (b22) {TEC};
	\node[rectangle, draw, semithick, right] at (6.9,3) (b23) {TEC};

	\draw[semithick] (l00.east) -- (l10.west);
	\draw[semithick] (l01.east) -- (l11.west);
	\draw[semithick,dashed] (l02.east) -- (l12.west);
	\draw[semithick,dashed] (l03.east) -- (l13.west);
	\draw[semithick,dashed] (l00.east) -- (l12.west);
	\draw[semithick,dashed] (l01.east) -- (l13.west);
	\draw[semithick] (l02.east) -- (l10.west);
	\draw[semithick] (l03.east) -- (l11.west);

	\draw[semithick,->] (l10.east) -- (b10.west);
	\draw[semithick,->] (l11.east) -- (b11.west);
	\draw[semithick,->] (l12.east) -- (b12.west);
	\draw[semithick,->] (l13.east) -- (b13.west);

	\draw[semithick] (b10.east) -- (l20.west);
	\draw[semithick,dashed] (b11.east) -- (l21.west);
	\draw[semithick] (b12.east) -- (l22.west);
	\draw[semithick,dashed] (b13.east) -- (l23.west);
	\draw[semithick,dashed] (b10.east) -- (l21.west);
	\draw[semithick] (b11.east) -- (l20.west);
	\draw[semithick,dashed] (b12.east) -- (l23.west);
	\draw[semithick] (b13.east) -- (l22.west);

	\draw[semithick,->] (l20.east) -- (b20.west);
	\draw[semithick,->] (l21.east) -- (b21.west);
	\draw[semithick,->] (l22.east) -- (b22.west);
	\draw[semithick,->] (l23.east) -- (b23.west);

	\draw[semithick,->] (b20.west) -- (b20.west);
	\draw[semithick,->] (b21.west) -- (b21.west);
	\draw[semithick,->] (b22.west) -- (b22.west);
	\draw[semithick,->] (b23.west) -- (b23.west);

\end{tikzpicture}
\caption{Synthetic channel construction for a polar code of length $N = 2^2 = 4$. Solid lines represent the $+$ transformation and dashed lines represent the $-$ transformation.}\label{fig:polartransformfaulty}
\end{figure}

\begin{figure*}
\footnotesize
\begin{align}
	E_{s,k,\delta}^{(\mathbf{s-})}	& = E_{s-1,k,\delta}^{(\mathbf{s})} + E_{s-1,k+2^{n-s},\delta}^{(\mathbf{s})} - E_{s-1,k,\delta}^{(\mathbf{s})}E_{s-1,k+2^{n-s},\delta}^{(\mathbf{s})} + \left(\overline{E_{s-1,k,\delta}^{(\mathbf{s})} + E_{s-1,k+2^{n-s},\delta}^{(\mathbf{s})} - E_{s-1,k,\delta}^{(\mathbf{s})}E_{s-1,k+2^{n-s},\delta}^{(\mathbf{s})}}\right)\Delta _{s,k}^{(\mathbf{s-})}, \label{eq:indup1faultyfull}\\
	E_{s,k,\delta}^{(\mathbf{s+})}	& = E_{s-1,k,\delta}^{(\mathbf{s})}E_{s-1,k+2^{n-s},\delta}^{(\mathbf{s})} + \left(\overline{E_{s-1,k,\delta}^{(\mathbf{s})}E_{s-1,k+2^{n-s},\delta}^{(\mathbf{s})}}\right)\Delta _{s,k}^{(\mathbf{s+})}. \label{eq:indup2faultyfull}
\end{align}
\hrule
\end{figure*}

\subsection{Fault Model}
We model faulty decoding as additional \emph{internal} erasures within the decoder, which may be caused either by faulty PEs or by faulty MEs (or both) and we assume, without loss of generality, that they manifest themselves when an output message is written to an ME. Moreover, we assume that these erasures are \emph{transient} in the sense that whenever an ME is written to, the internal erasures occur independently of any previous internal erasures. The partial sums, which are required by the $f^+$ update rule, also need to be stored in a memory, which however is typically smaller than the memory required to store the messages. Moreover, due to the partial sum recursive update rules~\cite[Proposition 3]{Arikan2009}, a single erasure in a partial sum will result in erasures in all following partial sums and we can intuitively see that the sensitivity of the SC decoder with respect to faults in the partial sum memory is high. Thus, in this work we assume that the partial sum memory is fault-free. 

Under the above assumptions, the internal erasures occur at the output of all synthetic channels of a polar code of blocklength $n$, i.e., $W_{s,k}^{(\mathbf{s})},~s = 1,\hdots,n,~\mathbf{s} \in \{+,-\}^s, ~k = 0,\hdots,2^{n-s}-1$. Moreover, the internal erasures occur independently of the message value and with probability $\delta$. Let us define a ternary-input erasure channel (TEC) with input alphabet $\mathcal{X} = \{{-1},0,{+1}\}$ and output alphabet $\mathcal{Y} = \mathcal{X}$ and the following transition probabilities
\begin{align}
	\Pr{0|0}	& = 1, \\
	\Pr{0|{-1}}	& = \Pr{0|{+1}} = \delta, \\
	\Pr{{+1}|{+1}}	& = \Pr{{-1}|{-1}} = 1-\delta,
\end{align}
where the probabilities of all remaining transitions are equal to zero. 

Using the above TEC, our error model can be represented as a cascade of a BEC\footnote{In order to avoid any confusion, we note that the erasure probability of this BEC corresponds to the expected erasure probability at a particular point within the \emph{deterministic} faulty-free decoder, where the expectation is taken over all possible noisy decoder input sequences. In other words, this BEC is not related to the randomness caused by the faulty decoder. The decoder noise is instead entirely modeled by the concatenated TEC.} with a TEC, as shown in Figure~\ref{fig:polartransformfaulty}, where $W_{s,k}^{(\mathbf{s})}$ results from the non-faulty polarizing channel transformation applied to a pair of channels $W_{s-1,k}^{(\mathbf{t})}$ and $W_{s-1,k+2^{n-s}}^{(\mathbf{t})}$ (where $\textbf{t}$ is a prefix of $\textbf{s}$) and ``TEC'' represents the internal erasures caused by the faulty SC decoder. We denote this cascaded compound channel by $W_{s,k,\delta}^{(\mathbf{s})}$ in order to make the dependence on $\delta$ explicit. It is easy to check that for $\delta = 0$ we get a non-faulty decoder, while for $\delta = 1$ all messages are always erasures leading to a fully faulty decoder. Since both of the aforementioned cases are already well understood, in the remainder of this paper we restrict $\delta$ to $\delta \in (0,1)$. 

In order to have a more rigorous definition of the internal erasure fault model, let us define the binary erasure indicator variable $\Delta _{s,k}^{(\mathbf{s})}$, where $\Delta _{s,k}^{(\mathbf{s})} = 1$ iff the TEC that comes after $W_{s,k}^{(\mathbf{s})}$ in Figure~\ref{fig:polartransformfaulty} causes an internal erasure at channel $W_{s,k}^{(\mathbf{s})}$, and $\Delta _{s,k}^{(\mathbf{s})} = 0$ otherwise. By definition, we have $\Pr{\Delta _{s,k}^{(\mathbf{s})}=1} = \delta$, thus $\Ex{\Delta _{s,k}^{(\mathbf{s})}} = \delta$ and $\text{var}\left[\Delta _{s,k}^{(\mathbf{s})}\right] = \delta(1-\delta)$. Since the internal erasures are assumed to be transient, all $\Delta _{s,k}^{(\mathbf{s})}$ are independent. Due to the cascaded BEC-TEC structure, we can rewrite \eqref{eq:indup1full} and \eqref{eq:indup2full} using $\Delta _{s,k}^{(\mathbf{s})}$ as \eqref{eq:indup1faultyfull} and \eqref{eq:indup2faultyfull}. In this case, for the binary erasure indicator variable $E_{s,k,\delta}^{(\mathbf{s})}$ we have $E_{s,k,\delta}^{(\mathbf{s})} = 1$ if and only if the output of the synthetic channel $W_{s,k,\delta}^{(\mathbf{s})}$ is an erasure and $E_{s,k,\delta}^{(\mathbf{s})} = 0$ otherwise. We note that, even though the special case of $E_{0,k,\delta}^{(\mathbf{s})}$ does not depend on $\delta$ but only on the erasure probability $p$ of the channel $W$, we keep the $\delta$ parameter in the notation for consistency.

Again, if we are only interested in the statistics of the indicator variable for a channel of a specific type $\mathbf{s}$, we can simplify~\eqref{eq:indup1faultyfull} and~\eqref{eq:indup2faultyfull} as
\begin{align}
	E_{s,\delta}^{(\mathbf{s-})}	& = {E_{s-1,\delta}^{(\mathbf{s})}}' + {E_{s-1,\delta}^{(\mathbf{s})}}'' - {E_{s-1,\delta}^{(\mathbf{s})}}'{E_{s-1,\delta}^{(\mathbf{s})}}'' \nonumber \\ 
																& + \left(\overline{{E_{s-1,\delta}^{(\mathbf{s})}}' + {E_{s-1,\delta}^{(\mathbf{s})}}'' - {E_{s-1,\delta}^{(\mathbf{s})}}'{E_{s-1,\delta}^{(\mathbf{s})}}''}\right)\Delta _{s}^{(\mathbf{s-})}, \label{eq:indup1faulty}\\
	E_{s,\delta}^{(\mathbf{s+})}	& = {E_{s-1,\delta}^{(\mathbf{s})}}'{E_{s-1,\delta}^{(\mathbf{s})}}'' + \left(\overline{{E_{s-1,\delta}^{(\mathbf{s})}}'{E_{s-1,\delta}^{(\mathbf{s})}}''}\right)\Delta _{s}^{(\mathbf{s+})}. \label{eq:indup2faulty}
\end{align}
where ${E_{s-1,\delta}^{(\mathbf{s})}}'$ and ${E_{s-1,\delta}^{(\mathbf{s})}}''$ denote two independent realizations of $E_{s-1,\delta}^{(\mathbf{s})}$, and $\Delta _{s}^{(\mathbf{s-})}$ and $\Delta _{s}^{(\mathbf{s+})}$ denote a realization of $\Delta _{s,k}^{(\mathbf{s-})}$ and $\Delta _{s,k}^{(\mathbf{s+})}$, respectively. The vector containing all $E^{(\mathbf{s})}_{s,\delta}$ indicator variables is denoted by $\mathbf{E}_{s,\delta}$.

We note that in a fully-parallel implementation, each ME has a corresponding PE, and our erasure-based fault model can take erasures in both the MEs and the PEs into account simultaneously. In a semi-parallel implementation, on the other hand, the MEs are significantly more than the PEs (i.e., typically $P \ll 2N-1$, as in~\cite{Leroux2013} where $N = 1024$ and $P = 64$), so it is reasonable to assume that faults stem only from the MEs, as the PEs can be made reliable with circuit-level techniques at a relatively low cost.

\subsection{Erasure Probability of Synthetic Channels Under Faulty SC Decoding}
Using the fault model introduced in the previous section, we can rewrite the recursive expressions for $Z_{s,k}^{(\mathbf{s})}$ (i.e., \eqref{eq:zup1full} and \eqref{eq:zup2full}) in order to obtain a recursive expression for the erasure probability of the synthetic channels in the faulty case, which we denote by $Z_{s,k,\delta}^{(\mathbf{s})} \triangleq \Ex{E_{s,k,\delta}^{(\mathbf{s})}}$. Specifically, we have
\begin{align}
	Z_{s,k,\delta}^{(\mathbf{s-})}	& = Z_{s-1,k,\delta}^{(\mathbf{s})} + Z_{s-1,k+2^{n-s},\delta}^{(\mathbf{s})} - Z_{s-1,k}^{(\mathbf{s})}Z_{s-1,k+2^{n-s},\delta}^{(\mathbf{s})} \nonumber \\ 
													& + \left(\overline{Z_{s-1,k,\delta}^{(\mathbf{s})} + Z_{s-1,k+2^{n-s},\delta}^{(\mathbf{s})} - Z_{s,k}^{(\mathbf{s})}Z_{s-1,k+2^{n-s},\delta}^{(\mathbf{s})}}\right)\delta, \label{eq:zup1faultyfull} \\
	Z_{s,k,\delta}^{(\mathbf{s+})}	& = Z_{s-1,k,\delta}^{(\mathbf{s})}Z_{s-1,k+2^{n-s},\delta}^{(\mathbf{s})} + \left(\overline{Z_{s-1,k,\delta}^{(\mathbf{s})}Z_{s-1,k+2^{n-s},\delta}^{(\mathbf{s})}}\right)\delta, \label{eq:zup2faultyfull}
\end{align}
with $Z_{0,k,\delta}^{(\emptyset)} = p,~k = 0,\hdots,2^n-1$. The channels $W^{(\mathbf{s})}_{s,k,\delta},~k = 0, \hdots, 2^{n-s}-1,$ are independent copies of the same type of channel, meaning that their erasure probabilities are identical. Thus, if we are only interested in the erasure probability of a specific \emph{type} $\mathbf{s}$ of channel we can simplify~\eqref{eq:zup1full} and~\eqref{eq:zup2full} by omitting the index $k$ as
\begin{align}
	Z_{s,\delta}^{(\mathbf{s}-)}	& = T_{\delta}^{-}\left(Z_{s-1,\delta}^{(\mathbf{s})}\right) \triangleq 2Z_{s-1,\delta}^{(\mathbf{s})} - \left(Z_{s-1,\delta}^{(\mathbf{s})}\right)^2 \nonumber \\
						& \qquad \qquad \qquad \quad + \left(\overline{2\Z{s-1}{s} - \left(\Z{s-1}{s}\right)^2}\right)\delta, \label{eq:zup1faulty} \\
	\Z{s}{s+}	& = T_{\delta}^{+}\left(\Z{s-1}{s}\right) \triangleq \left(\Z{s-1}{s}\right)^2 + \overline{\left(\Z{s-1}{s}\right)^2}\delta, \label{eq:zup2faulty}
\end{align}
with $Z_{0,\delta}^{(\emptyset)} = p$. The vector containing all $Z^{(\mathbf{s})}_{s,\delta},~\mathbf{s}\in\{+,-\}^s,$ variables is denoted by $\mathbf{Z}_{s,\delta}$. The random process $\epsilon _s$ can be rewritten for the faulty case as
\begin{align}
	\epsilon _{s,\delta} = \left\{ \begin{matrix} T_{\delta}^{+}(\epsilon_{s-1,\delta}) & \text{w.p. }~1/2, \\ T_{\delta}^{-}(\epsilon_{s-1,\delta}) & \text{w.p. }~1/2,  \end{matrix} \right. \label{eq:faulty}
\end{align}
with $\epsilon _{0,\delta} = Z(W) = p$.

\subsection{Properties of $T_{\delta}^+$ and $T_{\delta}^-$}
In this section, we show some properties of the $T^{+}_{\delta}$ and $T^{-}_{\delta}$ transformations, which will be useful to prove two negative results in the following section, as well as to interpret some of the numerical results of Section~\ref{sec:numerical}. 
\ifproofsend
	We note that the proofs of all properties can be found in the Appendix.
\fi
\begin{property}\label{property:geqdelta}
For $T^{+}_{\delta}(\epsilon)$ and $T^{-}_{\delta}(\epsilon)$, we have
\begin{enumerate}
	\setlength{\itemsep}{3pt}
	\item[(i)] $T^{+}_{\delta}(\epsilon) \geq \delta ,~\forall \epsilon, \delta \in \left[0,1\right]$,
	\item[(ii)] $T^{-}_{\delta}(\epsilon) \geq \delta ,~\forall \epsilon, \delta \in \left[0,1\right]$.
\end{enumerate}
\end{property}
\ifproofsend
	% do nothing
\else
\begin{IEEEproof}
For $T^{+}_{\delta}(\epsilon)$, we have
\begin{align}
	\epsilon ^2 + (1-\epsilon ^2)\delta & \geq \delta \Leftrightarrow \\
	(1-\delta)\epsilon ^2 & \geq 0,
\end{align}
which indeed holds for any $\epsilon, \delta \in \left[0,1\right]$. Similarly, for $T^{-}_{\delta}(\epsilon)$, we have
\begin{align}
	2\epsilon - \epsilon ^2 + (1-2\epsilon + \epsilon^2)\delta & \geq \delta \Leftrightarrow \\
	(1-\delta)(2\epsilon - \epsilon ^2) & \geq 0,
\end{align}
which indeed holds for any $\epsilon, \delta \in \left[0,1\right]$.
\end{IEEEproof}
\fi

\begin{property}\label{property:fixed}
The fixed points of $T^{+}_{\delta}(\epsilon)$ are $\epsilon = 1$ and $\epsilon = \frac{\delta}{1-\delta}$. The unique fixed point of $T^{-}_{\delta}(\epsilon)$ for $\epsilon \in [0,1]$ is $\epsilon = 1$.
\end{property}
\ifproofsend
	% do nothing
\else
\begin{IEEEproof}
The above property can easily be shown by solving $T^{+}_{\delta}(\epsilon) = \epsilon$ and $T^{-}_{\delta}(\epsilon) = \epsilon$ for $\epsilon$, respectively, and noting that one solution of $T^{-}_{\delta}(\epsilon) = \epsilon$ is negative.
\end{IEEEproof}
\fi
Moreover, the following two properties of the process $\epsilon _{s,\delta}$ give us some first insight into the effect that the faulty decoder has on the decoding process.
\begin{proposition}\label{proposition:submartingale}
The process $\epsilon _{s,\delta},~s = 0,1,\hdots,$ defined in \eqref{eq:faulty} is a submartingale.
\end{proposition}
\begin{IEEEproof}
Since $\epsilon _{s,\delta}$ is bounded, it holds that $\mathbb{E}(|\epsilon _{s,\delta}|) < \infty$. Moreover we have
\begin{align}
	\mathbb{E}(\epsilon _{s,\delta}|\epsilon _{s-1,\delta}) & = \frac{1}{2}\left(T^{+}_{\delta}(\epsilon_{s-1,\delta}) + T^{-}_{\delta}(\epsilon_{s-1,\delta})\right) \\
											& = \frac{1}{2}\left((1-\epsilon_{s-1,\delta} ^2)\delta + 2\epsilon_{s-1,\delta} \right. \nonumber \\
											& + \left. (1-2\epsilon_{s-1,\delta} + \epsilon_{s-1,\delta} ^2)\delta\right) \\
											& = \epsilon_{s-1,\delta} + (1-\epsilon_{s-1,\delta})\delta \geq \epsilon _{s-1,\delta}.
\end{align}
\end{IEEEproof}

\begin{proposition}\label{proposition:expclosedform}
For the expectation of the process $\epsilon _{s,\delta},~s = 0,1,\hdots,$ defined in \eqref{eq:faulty} we have
\begin{align}
	\mathbb{E}(\epsilon _{s,\delta}) & = 1- (1-p)(1-\delta)^s,
\end{align}
\end{proposition}
\begin{IEEEproof}
From the proof of Property~\ref{proposition:submartingale}, we know that 
\begin{align}
	\mathbb{E}(\epsilon _{s,\delta}|\epsilon _{s-1,\delta}) & = \epsilon_{s-1,\delta} + (1-\epsilon_{s-1,\delta})\delta. \label{eq:expej}
\end{align}
By taking the expectation with respect to $\epsilon _{s-1,\delta}$ on both sides of \eqref{eq:expej}, we have
\begin{align}
	\mathbb{E}(\epsilon _{s,\delta}) & = \mathbb{E}(\epsilon_{s-1,\delta}) + (1-\mathbb{E}((\epsilon_{s-1,\delta}))\delta \\
														& = (1-\delta)\mathbb{E}(\epsilon _{s-1,\delta}) + \delta, \label{eq:recurr}
\end{align}
with $\mathbb{E}(\epsilon _{0,\delta}) = \epsilon _{0,\delta} = p$. In order to simplify our notation for the proof, let $c_s \triangleq \mathbb{E}(\epsilon _{s,\delta})$. Then, \eqref{eq:recurr} can be written as
\begin{align}
	c_s 	& = (1-\delta)c_{s-1} + \delta.
\end{align}
By repeated substitution in the above expression we get
\begin{align}
	c_s 	& = (1-\delta)^2c_{s-2} + (1-\delta)\delta + \delta \\
			& = (1-\delta)^3c_{s-3} + (1-\delta)^2\delta + (1-\delta)\delta + \delta \\
			& = (1-\delta)^sc_{0} + \delta \sum _{n=0}^{s-1}(1-\delta)^{n}.
\end{align}
Since $c_{0} = p$ and $\sum _{n=0}^{s-1}(1-\delta)^{n} = \frac{1-(1-\delta)^s}{\delta}$, we finally have
\begin{align}
	\mathbb{E}(\epsilon _{s,\delta})& = (1-\delta)^sp + \delta \frac{1-(1-\delta)^s}{\delta} \\
									& = 1 - (1-p)(1-\delta)^s.
\end{align}
\end{IEEEproof}
Specifically, this tells us that, contrary to \cite[Section III-A]{Arikan2009}, the average erasure probability is not preserved by $T^{+}_{\delta}(\epsilon)$ and $T^{-}_{\delta}(\epsilon)$. Thus, even if fully reliable transmission were possible in the limit of infinite blocklength, the polar code would not be capacity achieving since $\lim _{s \rightarrow \infty}\Pr{\epsilon _{s,\delta} = 0} < 1-p$, meaning that the fraction of noiseless channels would be strictly smaller than the capacity of the BEC.

\subsection{Polarization Does Not Happen}
Unfortunately, as the following property shows, fully reliable transmission under faulty decoding is not possible.
\begin{property}\label{prop:nopol}
Let $\mathcal{Q}$ denote the sample space of the process $\epsilon_{s,\delta}$ and let $\epsilon_{s,\delta}(q),~q \in \mathcal{S},$ denote a specific realization of $\epsilon_{s,\delta}$ for $\delta > 0$. Polarization does not happen under faulty SC decoding for the BEC in the sense that $\nexists q \in \mathcal{Q}$ such that $\epsilon_{s,\delta}(q) \stackrel{s \rightarrow \infty}{\longrightarrow} 0$. 
\end{property}
\ifproofsend
	% do nothing
\else
\begin{IEEEproof}
This is a direct consequence of Property~\ref{property:geqdelta}, since all $\epsilon _{s,\delta}(q)$ are produced by repeated applications of $T^{+}_{\delta}$ and $T^{-}_{\delta}$ to $\epsilon _{0,\delta} = p$, so that $\epsilon _{s,\delta}(q) \geq \delta,~\forall q \in \mathcal{Q}$.
\end{IEEEproof}
\fi
It turns out that we can prove the following stronger result, which states that, under faulty SC decoding over the BEC, almost all channels become asymptotically useless.
\begin{proposition}\label{prop:as1}
For the process $\epsilon _{s,\delta},~s = 0,1,\hdots,$ defined in \eqref{eq:faulty} and for $\delta > 0$, we have $\epsilon _{s,\delta} \xrightarrow{\mathrm{a.s.}} 1$.
\end{proposition}

\begin{IEEEproof}
From Property~\ref{proposition:submartingale}, we know that $\epsilon _{s,\delta}$ is a bounded submartingale. Thus, it converges a.s. to some limiting random variable $\epsilon _{\infty}$. Moreover, from Proposition~\ref{proposition:expclosedform} we have 
\begin{align}
	\mathbb{E}(\epsilon _{s,\delta}) & = 1- (1-p)(1-\delta)^s,
\end{align}
which directly implies that $\lim _{s\rightarrow \infty}\mathbb{E}(\epsilon _{s,\delta}) = 1$, since, by assumption, $\delta \in (0,1)$. Equivalently, and since $\epsilon_{s,\delta} \in [0,1]$, we can write 
\begin{align}
	\lim _{s \rightarrow \infty}\mathbb{E}(|\epsilon _{s,\delta}-1|) & = 0,
\end{align}
which means, by definition, that $\epsilon _{s,\delta} \xrightarrow{L^1} 1$. Moreover, $\epsilon _{s,\delta} \xrightarrow{L^1} 1$ implies that $\epsilon _{s,\delta} \xrightarrow{\mathbb{P}} 1$. Since we know, due to the submartingale property, that $\epsilon _{s,\delta}$ also converges almost surely and almost sure convergence implies convergence in probability, all the aforementioned limits must be identical and we can conclude that $\epsilon _{s,\delta} \xrightarrow{\mathrm{a.s.}} 1$.
\end{IEEEproof}

\subsection{Synthetic Channel Ordering}
In the case of non-faulty decoding, there exists a partial ordering of the synthetic channels with respect to the BEC erasure probability $p$. In order to explain this ordering, we first need to define the notion of ``$\eta$-goodness''.
\begin{definition}
A synthetic channel $W_s^{(\mathbf{s})}$ is said to be ``$\eta$-good'' if $Z_{s}^{(\mathbf{s})} \leq \eta$.
\end{definition}
In the non-faulty case, it is easy to see that both $T^{+}(\epsilon)$ and $T^{-}(\epsilon)$ are increasing in $\epsilon,~\forall \epsilon \in [0,1]$. Thus, a synthetic channel that is $\eta$-good for a BEC with erasure probability $p_1$, will also be $\eta$-good for a BEC with erasure probability $p_2$ when $p_2 \leq p_1$.%\footnote{It must be emphasized that this does not necessarily mean that if a synthetic channel belongs to the information indices of a polar code constructed for BEC$(p_1)$, it will also belong to the set of information indices for a polar code constructed for BEC$(p_2)$, since we do not know how the remaining synthetic channels behave compared to the one in question. }

In this section, we show that under faulty decoding the partial ordering with respect to the BEC parameter $p$ is preserved and we show that a similar partial ordering exists with respect to the decoder erasure probability $\delta$. To this end, in the following two properties we examine the monotonicity of $T^{-}_{\delta}(\epsilon)$ and $T^{+}_{\delta}(\epsilon)$ with respect to $\epsilon$ and $\delta$.

\begin{property}\label{property:Tminusplusincreasing}
Both $T^{-}_{\delta}(\epsilon)$ and $T^{+}_{\delta}(\epsilon)$ are
\begin{enumerate}
	\setlength{\itemsep}{3pt}
	\item[(i)] Increasing in $\epsilon,~\forall \epsilon \in [0,1]$.
	\item[(ii)] Increasing in $\delta,~\forall \delta \in [0,1]$.
\end{enumerate}
\end{property}
\ifproofsend
	% do nothing
\else
\begin{IEEEproof}
(i) $T^{+}_{\delta}(\epsilon)$ can be re-written as
\begin{align}
	T^{+}_{\delta}(\epsilon) 	& = \epsilon ^2 + (1-\epsilon ^2)\delta \\
														& = \epsilon ^2(1-\delta) + \delta.
\end{align}
Thus, for any fixed $\delta \in [0,1]$, $T^{+}_{\delta}(\epsilon)$ is increasing in $\epsilon$ for any $\epsilon \in [0,1]$. Similarly, $T^{-}_{\delta}(\epsilon)$ can be re-written as
\begin{align}
	T^{-}_{\delta}(\epsilon) 	& = 2\epsilon - \epsilon ^2 + (1 - 2\epsilon + \epsilon ^2)\delta \\
														& = (2\epsilon - \epsilon ^2)(1-\delta) + \delta,
\end{align}
which is also increasing in $\epsilon$ for any $\epsilon \in [0,1]$.\\
(ii) Both $T^{-}_{\delta}(\epsilon)$ and $T^{+}_{\delta}(\epsilon)$ are linear functions of $\delta$ with a non-negative slope, so they are increasing $\forall \delta \in \mathbb{R}$.
\end{IEEEproof}
\fi

\begin{property}[Monotonicity with respect to $p$]\label{prop:monotonicityp}
Let $p_1,p_2 \in (0,1)$, $p_2 \leq p_1$ and $\delta \in (0,1)$. A synthetic channel that is $\eta$-good for a decoder with a fixed erasure probability $\delta$ over a BEC with erasure probability $p_1$ is also $\eta$-good for the same decoder over a BEC with erasure probability $p_2$.
\end{property}
\ifproofsend
	% do nothing
\else
\begin{IEEEproof}
The erasure probability of any synthetic channel $\W{s}{s}$ can be calculated by repeated applications of $T^{-}_{\delta}$ and $T^{+}_{\delta}$ starting from $p$ as
\begin{equation}
	\Z{s}{s}(p) = T^{s_s}_{\delta}\left(T^{s_{s-1}}_{\delta}\left(\cdots \left(T^{s_1}_{\delta}(p)\right) \right)\right),
\end{equation}
where $\mathbf{s} = [s_s,s_{s-1},\hdots,s_1]$ and $s_i \in \{+,-\}, i =1,\hdots,s$. Since from Property~\ref{property:Tminusplusincreasing}(i) we know that both $T^{-}_{\delta}(\epsilon)$ and $T^{+}_{\delta}(\epsilon)$ are increasing with respect to $\epsilon$, any composition of the two functions will also be increasing. Thus
\begin{equation}
	\Z{s}{s}(p_2) \leq \Z{s}{s}(p_1) \leq \eta.
\end{equation}
\end{IEEEproof}
\fi

%\subsubsection{Monotonicity With Respect to $\delta$}
The following proposition states that there also exists a partial ordering of the synthetic channels with respect to the decoder erasure probability $\delta$. This is a useful property, as it ensures that, for any given polar code, a decoder with internal erasure probability $\delta_2$ will not perform worse than a decoder with internal erasure probability $\delta_1$, where $\delta_2 \leq \delta_1$.
\begin{property}[Monotonicity with respect to $\delta$]\label{prop:monotonicityepsilon}
	Let $\delta_1,\delta_2 \in (0,1)$, $\delta_2 \leq \delta_1$ and $\epsilon \in (0,1)$. A synthetic channel that is $\eta$-good for a decoder with erasure probability $\delta _1$ over a BEC with a fixed erasure probability $\epsilon$ is also $\eta$-good for a decoder with erasure probability $\delta_2$ over the same channel.
\end{property}
\ifproofsend
	% do nothing
\else
\begin{IEEEproof}
Similarly to the proof of Property~\ref{prop:monotonicityp}, the proof stems directly from the monotonicity of $T^{-}_{\delta}(\epsilon)$ and $T^{+}_{\delta}(\epsilon)$ with respect to $\delta$ shown in Property~\ref{property:Tminusplusincreasing}(ii).
\end{IEEEproof}
\fi

\section{Frame Erasure Rate Under Faulty Decoding}\label{sec:fer}
In this section, we adapt the framework of~\cite{Bastani2013} to the case of faulty decoding in order to derive a lower bound on the frame erasure probability under faulty decoding. Let $P_e(\mathcal{A}_n)$ denote the frame erasure rate (FER) of a polar code of length $2^n$ with information set $\mathcal{A}_n$. From \cite[Section V-B]{Arikan2009}, we have the general upper bound
\begin{align}
	P_e(\mathcal{A}_n) & \leq \sum_{\mathbf{s} \in \mathcal{A}_n} \Znf{n}{s} \triangleq P_{e}^{\mathrm{UB}}. \label{eqn:ub}
\end{align}
Furthermore, from \cite{Bastani2013} we have the lower bound
\begin{align}
	P_e(\mathcal{A}_n) 	& \geq \sum_{\mathbf{s} \in \mathcal{A}_n} \Znf{n}{s} - \frac{1}{2}\sum_{\substack{\mathbf{s},\mathbf{t} \in \mathcal{A}_n:\\\mathbf{s} \neq \mathbf{t}}}\left(\Znf{n}{s}\Znf{n}{t} + \Cnf{n}{s}{t} \right) \triangleq P_{e}^{\mathrm{LB}} \label{eqn:lb}
\end{align}
where $\mathbf{C}_n \triangleq [C_{n}^{(\mathbf{s},\mathbf{t})}: \mathbf{s},\mathbf{t} \in \{+,-\}^n]$ denotes the covariance matrix of the random vector $\mathbf{E}_{n}$, where $C^{(\mathbf{s},\mathbf{t})}_{n} \triangleq \mathrm{cov}[\Enf{n}{\mathbf{s}}\Enf{n}{\mathbf{t}}]$. It was shown in~\cite{Bastani2013} that, in the non-faulty case, the elements of $\mathbf{C}_s, s = 1,\hdots,n$, can be calculated recursively from the elements of $\mathbf{C}_{s-1}$ and $Z_{s-1}^{(\mathbf{s})}$ as follows
\begin{align}
	\Cnf{s}{s-}{t-}	& = 2\overline{\Znf{s-1}{s}\Znf{s-1}{t}}\Cnf{s-1}{s}{t} + {\Cnf{s-1}{s}{t}}^2, \label{eqn:Cn--nf} \\
	\Cnf{s}{s-}{t+}	& = 2\overline{\Znf{s-1}{s}}\Znf{s-1}{t}\Cnf{s-1}{s}{t} - {\Cnf{s-1}{s}{t}}^2, \label{eqn:Cn-+nf} \\
	\Cnf{s}{s+}{t-}	& = 2\Znf{s-1}{s}\overline{\Znf{s-1}{t}}\Cnf{s-1}{s}{t} - {\Cnf{s-1}{s}{t} }^2, \label{eqn:Cn+-nf} \\
	\Cnf{s}{s+}{t+}& =  2\Znf{s-1}{s}\Znf{s-1}{t}\Cnf{s-1}{s}{t}  + {\Cnf{s-1}{s}{t} }^2, \label{eqn:Cn++nf}
\end{align}
with $\Cnf{0}{\emptyset}{\emptyset} = p(1-p)$. In the case of reliable decoding, the second sum in~\eqref{eqn:lb} goes to zero as $n$ is increased~\cite{Bastani2013} if $R = \frac{|\mathcal{A}_n|}{2^n} < 1 - p$, so that
\begin{align}
	P_e(\mathcal{A}_n) \approx \sum_{\mathbf{s} \in \mathcal{A}_n} \Znf{n}{s}. \label{eqn:FER}
\end{align}
We can use the upper and lower bounds of~\eqref{eqn:lb} and~\eqref{eqn:ub} for the case of faulty decoding by replacing $\Znf{n}{s}$ with $\Z{n}{s}$, and $\Cnf{n}{s}{t}$ with $\C{n}{s}{t}$, where $\mathbf{C}_{n,\delta}^{(\mathbf{s},\mathbf{t})} \triangleq [\C{n}{s}{t}: \mathbf{s},\mathbf{t} \in \{+,-\}^n],$ is the covariance matrix of the random vector $\mathbf{E}_{n,\delta}$. In the case of faulty decoding, as $n$ is increased, we know from Proposition~\ref{prop:as1} that almost all $\Z{n}{s}\Z{n}{t},~\mathbf{s},\mathbf{t}\in \mathcal{A}_n,$ are equal to $1$. Moreover, the non-diagonal elements of $\C{n}{s}{t}$ still converge to $0$ for any $\mathbf{s},\mathbf{t}$, as almost all indicator variables become deterministic like in the fault-free case. Thus, for some $n$ the lower bound of~\eqref{eqn:lb} becomes negative and can be replaced by the trivial lower bound $P_e(\mathcal{A}_n) \geq \max_{\mathbf{s} \in \mathcal{A}_n} \Z{n}{s}$. Similarly, for some $n$ the upper bound of~\eqref{eqn:ub} becomes greater than $1$, so it can be replaced by the trivial upper bound $P_e(\mathcal{A}_n) \leq 1$. Clearly though, since $\Z{n}{s}$ converges to $1$ as $n$ grows for almost all $\mathbf{s} \in \{+,-\}^n,$ we have $\lim _{n \rightarrow \infty}P_e(\mathcal{A}_n) = 1$ for any $\mathcal{A}_n$ such that $\lim _{n \rightarrow \infty} \frac{|\mathcal{A}_n|}{2^n} \nrightarrow 0$.

\subsection{Lower Bound on $P_e(\mathcal{A}_n)$ Under Faulty Decoding}
We already have an efficient way to calculate $\Z{n}{s}$ recursively (i.e., \eqref{eq:zup1faulty} and \eqref{eq:zup2faulty}), but, in order to evaluate $P_{e}^{\mathrm{LB}}$, we still need to find an efficient way to calculate $\mathbf{C}_{n,\delta}$. To this end, we first introduce a property which we then combine with the results of~\cite{Bastani2013} in order to obtain a recursive expression for $\mathbf{C}_{s,\delta},~s = 1,\hdots,n$.
\begin{property}\label{property:covariance}
Let $X,Y$ denote two arbitrary random variables. Let $\Delta_1, \Delta_2$ denote two random variables with $\Delta_1, \Delta_2 \in \{0,1\}$ and $\mathbb{E}\left[\Delta_1\right] = \mathbb{E}\left[\Delta_2\right] = \delta$ that are independent of $X,Y$ and of each other. Then, we have
\begin{align}
	\mathrm{cov}\left[X + (1-X)\Delta_1,Y + (1-Y)\Delta_2\right]	& = (1-\delta)^2\mathrm{cov}\left[X,Y\right].
\end{align}
\end{property}
\ifproofsend
	% do nothing
\else
\begin{IEEEproof}
For simpler notation, let us define $X' \triangleq X + (1-X)\Delta_1$ and $Y' \triangleq Y + (1-Y)\Delta_2$. We then have
\begin{align}
	\mathrm{cov}\left[X',Y'\right]	& = \mathbb{E}[X'Y'] - \mathbb{E}[X']\mathbb{E}[Y'] \\
									& = \mathbb{E}[(1-\Delta_1)X + \Delta_1)((1-\Delta_2)Y + \Delta_2)] \nonumber \\
									& - \mathbb{E}[(1-\Delta_1)X + \Delta_1]\mathbb{E}[(1-\Delta_2)Y + \Delta_2] \\
									& \stackrel{(*)}{=} \Ex{(1-\Delta_1)(1-\Delta_2)}(\mathbb{E}[XY] - \mathbb{E}[X]\mathbb{E}[Y]) \\
									& \stackrel{(**)}{=} (1-\delta)^2\mathrm{cov}\left[X,Y\right],
\end{align}
where for $(*)$ we have used the independence of $\Delta_1$ and $\Delta_2$ from $X$ and $Y$, while for $(**)$ we have used the independence between $\Delta_1$ and $\Delta_2$.
\end{IEEEproof}
\fi

\begin{proposition}
The covariance matrix of the random vector $\mathbf{E}_{s,\delta}$, denoted by $\mathbf{C}_{s,\delta} \triangleq [C_{s,\delta}^{(\mathbf{s},\mathbf{t})}: \mathbf{s},\mathbf{t} \in \{+,-\}^s]$, where $\mathbf{C}_{s,\delta} \triangleq \mathrm{cov}\left[\E{s}{\mathbf{s}}\E{s}{\mathbf{t}}\right]$, can be computed in terms of $\mathbf{C}_{s-1,\delta}$ and $\mathbf{Z}_{s-1,\delta}$ as follows:
\begin{align}
	\C{s}{s-}{t-}	& = \left(1-\delta\right)^2\left(2\overline{\Z{s-1}{s}\Z{s-1}{t}}\C{s-1}{s}{t} + {\C{s-1}{s}{t}}^2\right), \label{eqn:Cn--} \\
	\C{s}{s-}{t+}	& = \left(1-\delta\right)^2\left(2\overline{\Z{s-1}{s}}\Z{s-1}{t}\C{s-1}{s}{t} - {\C{s-1}{s}{t}}^2\right), \label{eqn:Cn-+} \\
	\C{s}{s+}{t-}	& = \left(1-\delta\right)^2\left(2\Z{s-1}{s}\overline{\Z{s-1}{t}}\C{s-1}{s}{t} - {\C{s-1}{s}{t} }^2\right), \label{eqn:Cn+-} \\
	\C{s}{s+}{t+}& = \left(1-\delta\right)^2\left(2\Z{s-1}{s}\Z{s-1}{t}\C{s-1}{s}{t}  + {\C{s-1}{s}{t} }^2\right), \label{eqn:Cn++}
\end{align}
with $\Cnf{0}{\emptyset}{\emptyset} = p(1-p)$.
\end{proposition}
\begin{IEEEproof}
To avoid unnecessary repetition, we prove the result only for \eqref{eqn:Cn++}, as the remaining relations \eqref{eqn:Cn--}--\eqref{eqn:Cn+-} can be derived in the same way. Recall that, in the case of faulty decoding, from~\eqref{eq:indup2faulty} we have
\begin{align}
	E_{s,\delta}^{(\mathbf{s+})}	& = {E_{s-1,\delta}^{(\mathbf{s})}}'{E_{s-1,\delta}^{(\mathbf{s})}}'' + \left(1-{E_{s-1,\delta}^{(\mathbf{s})}}'{E_{s-1,\delta}^{(\mathbf{s})}}''\right)\Delta _{s}^{(\mathbf{s+})}, \label{eqn:s+} \\
	E_{s,\delta}^{(\mathbf{t+})}	& = {E_{s-1,\delta}^{(\mathbf{t})}}'{E_{s-1,\delta}^{(\mathbf{t})}}'' + \left(1-{E_{s-1,\delta}^{(\mathbf{t})}}'{E_{s-1,\delta}^{(\mathbf{t})}}''\right)\Delta _{s}^{(\mathbf{t+})}. \label{eqn:t+}
\end{align}
Let us define $X \triangleq {\E{s-1}{\mathbf{s}}}'{\E{s-1}{\mathbf{s}}}''$, $Y \triangleq {\E{s-1}{\mathbf{t}}}'{\E{s-1}{\mathbf{t}}}''$, $\D{s}{s+} \triangleq \Delta_1$, and $\D{s}{t+} \triangleq \Delta_2$. Then, we can rewrite \eqref{eqn:s+} as
\begin{align}
	\E{n}{\mathbf{s+}} & = X + (1-X)\Delta_1, \label{eqn:s+2} \\
	\E{n}{\mathbf{t+}} & = Y + (1-Y)\Delta_2, \label{eqn:t+2}
\end{align}
where $X$ and $Y$ are identical to the update rule for $\Enf{s}{\mathbf{s+}}$ and $\Enf{s}{\mathbf{t+}}$ in the fault-free case given in \eqref{eq:indup2}, respectively. Using $\mathbb{E}\left[\D{s}{s+}\right] = \mathbb{E}\left[\D{s}{t+}\right] = \delta,$ along with the fact that $\D{s}{s+}$ and $\D{s}{t+}$ are independent by assumption, we can apply Proposition~\ref{property:covariance} to the update formula for $\mathrm{cov}\left[X,Y\right]$ from \cite{Bastani2013} given in \eqref{eqn:Cn++nf}, in order to obtain \eqref{eqn:Cn++}.
\end{IEEEproof}
It is intuitively pleasing to note that, for $\delta = 0$ (i.e., for fault-free decoding), the expressions in~\eqref{eqn:Cn--}--\eqref{eqn:Cn++} become identical to the expressions in~\eqref{eqn:Cn--nf}--\eqref{eqn:Cn++nf}.

\section{Unequal Error Protection}\label{sec:protection}
As mentioned in Section~\ref{sec:introduction}, standard methods employed to enhance the fault tolerance of circuits, such as using larger transistors or circuit-level error correcting codes, are costly in terms of both area and power if the entire circuit needs to be protected. With this in mind, we note that in SC decoding of polar codes not all levels in the tree of PEs are of equal importance, meaning that it may suffice to employ \emph{partial protection} of the decoder against hardware-induced errors. In fact, we shall see in Proposition~\ref{proposition:uneqprot}, a careful application of such a protection method allows polarization to happen even in a faulty decoder while protecting only a constant fraction of the total decoder PEs. The concept of identifying and protecting the most critical part of a decoder has also been used in the literature related to faulty decoding of LDPC codes. For example, in~\cite{Ngassa2014,KameniNgassa2015} it is found that a noiseless implementation of the early-termination circuitry can significantly improve the error-correcting performance of a noisy LDPC decoder.

%\subsection{Partial Stage Protection}
Let $0 \leq n_\text{p} \leq n+1$ denote the number of levels that are protected, starting from level $n$ of the tree (i.e., the root) and going towards level $0$ of the tree (i.e., the leaves). We assume that for these $n_\text{p}$ levels we have $\delta = 0$, meaning that $n_\text{p} = n+1$ results in a fault-free decoder and $n_\text{p} = 0$ is equivalent to the faulty SC decoder defined in Section~\ref{sec:faulty}. Let $N_{\text{p}}$ denote the total number of protected PEs, where 
\begin{align}
	N_{\text{p}} = \left\{ \begin{matrix} \sum _{j=0}^{n_{\text{p}}-1}2^{j} = 2^{n_{\text{p}}}-1, & n_{\text{p}} > 0,\\ 0, & n_{\text{p}} = 0. \end{matrix} \right.
\end{align}
If we set $n_{\text{p}} = (n+1) - n_{\text{u}}$, where $n_{\text{u}} > 0$ is a \emph{fixed} number of unprotected levels, then the fraction of the decoder that is protected converges to a constant as $n$ grows. Indeed, we have
\begin{align}
	\lim _{n \rightarrow \infty}\frac{N_{\text{p}}}{N_{\text{PE}}} & = \lim _{n \rightarrow \infty}\frac{2^{(n + 1) - n_{\text{u}}}-1}{2^{n+1}-1} = 2^{-n_{\text{u}}}.
\end{align}
In this case, the process $\epsilon_{s,\delta}$ can be rewritten as
\begin{align}
	\epsilon _{s,\delta} = \left\{ \begin{matrix} 
															T_{\delta}^{+}(\epsilon_{s-1,\delta}), & \text{w.p. } 1/2, & \multirow{2}{*}{if $s = 1,\hdots,n_{\text{u}}$,} \\ 
															T_{\delta}^{-}(\epsilon_{s-1,\delta}), & \text{w.p. } 1/2, &  \\
															T^{+}(\epsilon_{s-1,\delta}), & \text{w.p. } 1/2, & \multirow{2}{*}{if $s = n_{\text{u}}+1,\hdots,n$.} \\
															T^{-}(\epsilon_{s-1,\delta}), & \text{w.p. } 1/2, & 
														\end{matrix} \right. \label{eq:protej}
\end{align}
The following proposition asserts that the protection of a constant fraction of the decoder is sufficient to ensure that polarization happens as $n$ grows.

\begin{proposition}\label{proposition:uneqprot}
Setting $n_{\text{p}} = s - n_{\text{u}}$ for any fixed $n_{\text{u}}$ suffices to ensure that $\epsilon _{s,\delta}$ converges almost surely to a random variable $\epsilon _{\infty} \in \{0,1\}$. However, the unprotected levels result in a rate loss $\Delta R (\delta,p,n_{\text{u}})$, in the sense that $P(\epsilon _{\infty} = 0) = 1 - p - \Delta R (\delta,p,n_{\text{u}})$, which can be calculated in closed form as
\begin{align}
	\Delta R (\delta,p,n_{\text{u}})	& = (1-(1-\delta)^{n_{\text{u}}})(1-p).
\end{align}
\end{proposition}
\begin{IEEEproof}
The process $\epsilon_{s,\delta}$ as defined in \eqref{eq:protej} is a submartingale for $s \leq n_{\text{u}}$, but it becomes a martingale for $s > n_{\text{u}}$. Thus, for $s > n_{\text{u}}$ we have $\mathbb{E}(\epsilon _{s,\delta}) = \mathbb{E}(\epsilon _{n_{\text{u}},\delta})$. Using the arguments from \cite[Proposition 9]{Arikan2009}, we can show that $\epsilon _{s,\delta}$ converges almost surely to a random variable $\epsilon _{\infty} \in \{0,1\}$ with $P(\epsilon _{\infty} = 0) = 1-\mathbb{E}(\epsilon _{n_{\text{u}}}) \leq 1 - p$. Equivalently, $P(\epsilon _{\infty} = 0) = 1 - p - \Delta R (\delta,\epsilon,n_{\text{u}})$ for $\Delta R (\delta,\epsilon,n_{\text{u}}) = \mathbb{E}(\epsilon_{n_{\text{u}}}) - p$. Using the closed form expression for $\mathbb{E}(\epsilon_{s,\delta})$ from Proposition~\ref{proposition:expclosedform}, we get
\begin{align}
	\Delta R (\delta,p,n_{\text{u}}) & = \mathbb{E}(\epsilon_{n_{\text{u}}}) - p \\
																					& = 1-(1-p)(1-\delta)^{n_{\text{u}}} - p \\
																					& = \left(1-\left(1-\delta\right)^{n_{\text{u}}}\right)(1-p).
\end{align}
\end{IEEEproof}

Proposition~\ref{proposition:uneqprot} implies that, when partial protection of the decoder is employed, polar codes are still not capacity achieving, but they can nevertheless be used for reliable transmission at any rate $R$ such that $R < 1-p-\Delta R (\delta,p,n_{\text{u}})$. 

\section{Optimal Blocklength Under Faulty Decoding}\label{sec:optbl}
In the finite blocklength regime, which is of practical interest, there are two clashing effects occurring. On one side, we have the polarization process, which tends to \emph{decrease} the code's FER as the blocklength is increased, but on the other side we have the internal erasures of the decoder which tend to \emph{increase} the code's FER as the blocklength is increased. From Proposition~\ref{prop:as1} we already know that, as the blocklength is increased towards infinity, the latter effect dominates and the resulting polar code becomes asymptotically useless. However, there must exist at least one blocklength which minimizes the FER and it is of great practical interest to identify this length.

Since this is a finite-length problem with practical applications, there will usually be a pre-defined maximum blocklength $n_{\max}$ for which a decoder is implementable with acceptable complexity. Thus, for a given $n_{\max}$, we define $\mathcal{N} = \{0,\hdots,n_{\max}\}$ as the set of $n$ values of interest.  For a given code rate $R$, we define the $n^*$ which leads to the optimal blocklength $N^* = 2^{n^*}$ as
\begin{align}
	n^* & = \arg \min _{n \in \mathcal{N}} P_e(\mathcal{A}_n). \label{eqn:optn}
\end{align}

A simple way to identify the optimal blocklength is to perform extensive Monte-Carlo simulations of the codes for all $n \in \mathcal{N}$. However, we can find the solution more efficiently by using the bounds on $P_{e}(\mathcal{A}_n)$ given by~\eqref{eqn:ub} and~\eqref{eqn:lb}. First, we study the special case where $p < \delta$. More specifically, the following proposition shows that, when $p < \delta$, it is optimal in terms of the FER to use \emph{uncoded} transmission, as the faulty decoder can only increase the FER.
\begin{proposition}
If $p < \delta$, then $n^* = 0$.
\end{proposition}
\begin{IEEEproof}
The FER for $n = 0$ (i.e., uncoded transmission) over a BEC$(p)$ is equal to $p$. From Property~\ref{property:geqdelta}, we know that $\Z{n}{s} \geq \delta,~\forall \mathbf{s}\in\{+,-\}^n$. Since $p < \delta$ by assumption, we have $\Z{n}{s} > p,~\forall \mathbf{s}\in\{+,-\}^n$. Thus, using the trivial lower bound on the FER, i.e., $P^{\mathrm{LB}}_e = \max _{\mathbf{s} \in \mathcal{A}_n}\Z{n}{s}$, we can see that $P^{\mathrm{LB}}_e > p$ for any $\mathcal{A}_n$ such that $|\mathcal{A}_n| > 0$. Thus, in this special case coded transmission with any blocklength such that $n > 0$ and at any rate $R > 0$, leads to a higher FER than uncoded transmission.
\end{IEEEproof}
In general, we can efficiently evaluate $P_{e}^{\mathrm{UB}}(\mathcal{A}_n)$ and $P_{e}^{\mathrm{LB}}(\mathcal{A}_n)$ for all $n \in \mathcal{N}$ for a given rate $R$. Using these values, we can deduce whether there exists a single $n \in \mathcal{N}$ satisfying the following inequality
\begin{align}
	P_{e}^{\mathrm{UB}}(\mathcal{A}_n) \leq P_{e}^{\mathrm{LB}}(\mathcal{A}_{n'}),~\forall n' \in \mathcal{N}. \label{eqn:singlen}
\end{align}
If there exists such a unique $n \in \mathcal{N}$, then clearly this is the optimal $n^*$. Otherwise, we need to examine (via Monte-Carlo simulations) all $n \in \mathcal{N}$ for which $P_{e}^{\mathrm{UB}}(\mathcal{A}_n)$ and $P_{e}^{\mathrm{LB}}(\mathcal{A}_n)$ overlap, i.e., for which $\exists n' \in \mathcal{N}$ and $\exists \mathrm{B} \in \{\mathrm{UB},\mathrm{LB}\}$ such that
\begin{align}
	P_{e}^{\mathrm{LB}}(\mathcal{A}_{n'}) & \leq P_{e}^{\mathrm{B}}(\mathcal{A}_n) \leq P_{e}^{\mathrm{UB}}(\mathcal{A}_{n'}).
\end{align}
Numerical results for $n^*$ using the above observations are presented in Section~\ref{sec:optblock}.

\section{Numerical results}\label{sec:numerical}
In this section we provide some numerical results to explore the process $\epsilon _{s,\delta}$, as well as the FER performance of polar codes constructed based on this process. Moreover, we use the FER bounds derived in Section~\ref{sec:fer} in order to find the optimal blocklength for a polar code under faulty SC decoding and we explore the effectiveness of the unequal error protection scheme described in Section~\ref{sec:protection}. 

\paragraph*{Remark} We note that most of the results in this section are presented for a decoder erasure probability of $\delta = 10^{-6}$. From Property~\ref{property:geqdelta}, we know that the erasure probability of the synthetic channels is lower bounded by $\delta$. Moreover, from \eqref{eqn:ub}, we know that the frame error rate is upper bounded by the sum of the erasure probabilities of the synthetic channels used to transmit information. In the numerical experiments we did, we saw that the same number also provides a good lower bound for most code rates. Moreover, the first numerical results are also provided for $\delta = 10^{-4}$ and they show that the behavior of the decoder does not seem to change fundamentally for different values of $\delta$. Thus, have we selected $\delta = 10^{-6}$ as this leads to frame error rates that are practically relevant for the blocklengths that we have considered. 

\subsection{Bhattacharyya Parameters $\Z{n}{s}$}\label{sec:numericalpolarization}
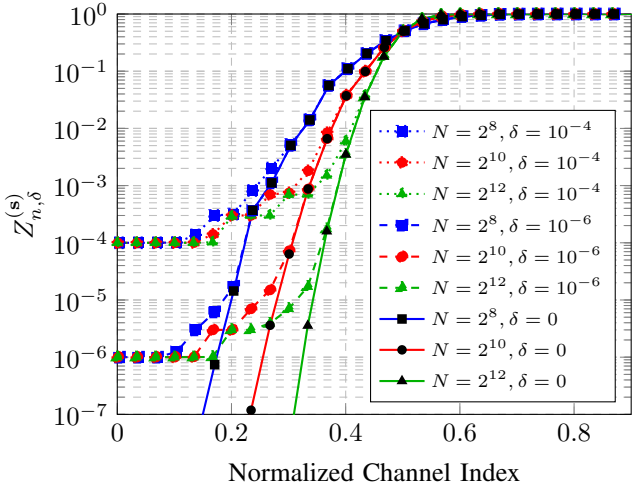
\begin{figure}[t]
	\centering
	\begin{tikzpicture}

	\pgfplotsset{grid style={dashed}}

	\begin{semilogyaxis}[
		width = \figurewidth,
		height = \figureheight,
		xlabel = {Normalized Channel Index},
		ylabel = {$\Z{n}{s}$},
		xmin = 0, xmax = 0.9,
		restrict x to domain=0:1,
		ymin = 1e-7, ymax = 1,
		grid = both,
		legend style={legend pos=south east,font=\scriptsize},
		legend cell align=left,
		% Use '\\' delimiter because comma in legend entry causes issues
		legend entries={$N = 2^8, \delta = 10^{-4}$\\ $N = 2^{10}, \delta = 10^{-4}$\\ $N = 2^{12}, \delta = 10^{-4}$\\ $N = 2^8, \delta = 10^{-6}$\\ $N = 2^{10}, \delta = 10^{-6}$\\ $N = 2^{12}, \delta = 10^{-6}$\\ $N = 2^8, \delta = 0$\\ $N = 2^{10}, \delta = 0$\\ $N = 2^{12}, \delta = 0$\\},
	]		

		% faulty delta = 1e-04
		\addplot[blue, thick, dotted, mark=square*] table {fig/data/Z_n8_faulty_1e-04.dat};
		\addplot[red, thick, dotted, mark=*] table {fig/data/Z_n10_faulty_1e-04.dat};
		\addplot[darkgreen, thick, dotted, mark=triangle*, mark options={scale=1.2}] table {fig/data/Z_n12_faulty_1e-04.dat};
	
		% faulty delta = 1e-06
		\addplot[blue, thick, dashed, mark=square*] table {fig/data/Z_n8_faulty_1e-06.dat};
		\addplot[red, thick, dashed, mark=*] table {fig/data/Z_n10_faulty_1e-06.dat};
		\addplot[darkgreen, thick, dashed, mark=triangle*, mark options={scale=1.2}] table {fig/data/Z_n12_faulty_1e-06.dat};
	
		% non-faulty
		\addplot[blue, thick, solid, mark=square*, mark options={solid,black,scale=0.7}] table {fig/data/Z_n8_nonFaulty.dat};
		\addplot[red, thick, solid, mark=*, mark options={solid,black,scale=0.7}] table {fig/data/Z_n10_nonFaulty.dat};
		\addplot[darkgreen, thick, solid, mark=triangle*, mark options={solid,black,scale=0.8}] table {fig/data/Z_n12_nonFaulty.dat};

	\end{semilogyaxis}

\end{tikzpicture}%
	\caption{Sorted $\Z{n}{s},~\mathbf{s} \in \{+,-\}^n$ and $\Znf{n}{s},~\mathbf{s} \in \{+,-\}^n$, values for polar codes of length $N = 256,1024,4096,$ designed for the BEC$(0.5)$ under faulty SC decoding.}\label{fig:Z}
\end{figure}

In Figure~\ref{fig:Z}, we show the sorted values $\Z{n}{s},~\mathbf{s} \in \{+,-\}^n$, for polar codes with $n = 8,10,12,$ designed for the BEC$(0.5)$ under faulty SC decoding with $\delta = 10^{-4}, \delta = 10^{-6}$,  and $\delta = 0$. We observe that we always have $\Z{n}{s} \geq \delta$, as predicted by Property~\ref{property:geqdelta}. Moreover, $\epsilon = \frac{\delta}{1-\delta}$ is a fixed point of $T^{+}_{\delta}(\epsilon)$, but it is not a fixed point of $T^{-}_{\delta}(\epsilon)$ (whereas $\epsilon = 1$ is a fixed point for both), resulting in the staircase-like structure that we can observe in Figure~\ref{fig:Z}. Finally, we see that the behavior of the faulty decoder does not change fundamentally when increasing the value of $\delta$.

\subsection{Frame Erasure Rate}
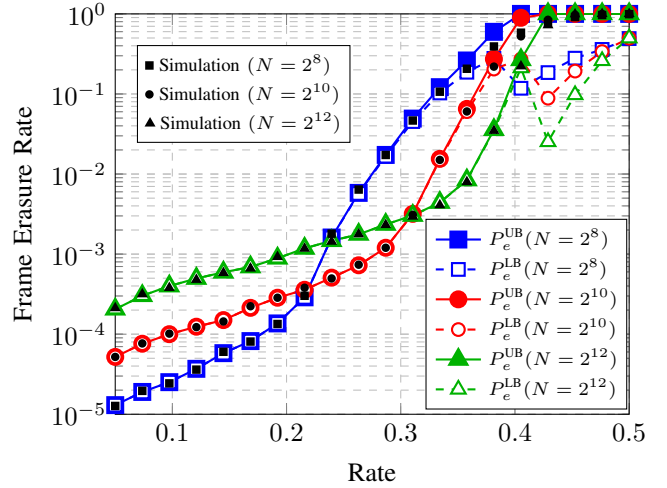
\begin{figure}[t]
	\centering
	\begin{tikzpicture}

	\pgfplotsset{grid style={dashed}}

	\begin{semilogyaxis}[
		width = \figurewidth,
		height = \figureheight,
		xlabel = {Rate},
		ylabel = {Frame Erasure Rate},
		xmin = 0.05, xmax = 0.5,
		ymin = 1e-5, ymax = 1,
		grid = both,
		%legend style={legend pos=south east,font=\scriptsize},
		%legend cell align=left,
		%legend entries={$P_e^{\text{UB}} (N = 2^8)$, $P_e^{\text{LB}} (N = 2^8)$, $P_e^{\text{UB}} (N = 2^{10})$, $P_e^{\text{LB}} (N = 2^{10})$, $P_e^{\text{UB}} (N = 2^{12})$, $P_e^{\text{LB}} (N = 2^{12})$},
	]		

		% n = 8
		\addplot[blue, thick, solid, mark=square*, mark options={scale=1.5}] table[x index=0, y index = 1] {fig/data/UB_LB_n8.dat};
		\label{n8_UB}
		\addplot[blue, thick, dashed, mark=square*, mark options={solid,blue,fill=white,scale=1.2}] table[x index=0, y index = 2] {fig/data/UB_LB_n8.dat};
		\label{n8_LB}

		% n = 10
		\addplot[red, thick, solid, mark=*, mark options={scale=1.5}] table[x index=0, y index = 1] {fig/data/UB_LB_n10.dat};
		\label{n10_UB}
		\addplot[red, thick, dashed, mark=*, mark options={solid,red,fill=white,scale=1.2}] table[x index=0, y index = 2] {fig/data/UB_LB_n10.dat};
		\label{n10_LB}

		% n = 12
		\addplot[darkgreen, thick, solid, mark=triangle*, mark options={scale=2}] table[x index=0, y index = 1] {fig/data/UB_LB_n12.dat};
		\label{n12_UB}
		\addplot[darkgreen, thick, dashed, mark=triangle*, mark options={solid,darkgreen,fill=white,scale=1.6}] table[x index=0, y index = 2] {fig/data/UB_LB_n12.dat};
		\label{n12_LB}
		
		% Simulations
		\addplot[only marks, blue, thick, dotted, mark=square*, mark options={solid,black,fill=black,scale=0.6}] table[x index=0, y index = 1] {fig/data/MC_N256_1e-6.dat};
		\label{n8_MC}
		\addplot[only marks, red, thick, dotted, mark=*, mark options={solid,black,fill=black,scale=0.6}] table[x index=0, y index = 1] {fig/data/MC_N1024_1e-6.dat};
		\label{n10_MC}
		\addplot[only marks, darkgreen, thick, dotted, mark=triangle*, mark options={solid,black,fill=black,scale=0.8}] table[x index=0, y index = 1] {fig/data/MC_N4096_1e-6.dat};
		\label{n12_MC}
		
		\node [draw,fill=white] at (rel axis cs: 0.8,0.25) {\scriptsize \shortstack[l]{
		\ref{n8_UB} $P_e^{\text{UB}} (N = 2^8)$ \\
		\ref{n8_LB} $P_e^{\text{LB}} (N = 2^8)$ \\
		\ref{n10_UB} $P_e^{\text{UB}} (N = 2^{10})$ \\
		\ref{n10_LB} $P_e^{\text{LB}} (N = 2^{10})$ \\
		\ref{n12_UB} $P_e^{\text{UB}} (N = 2^{12})$ \\
		\ref{n12_LB} $P_e^{\text{LB}} (N = 2^{12})$
		}
		};
		
		\node [draw,fill=white] at (rel axis cs: 0.25,0.8) {\scriptsize \shortstack[l]{
		\ref{n8_MC} Simulation $(N = 2^8)$ \\
		\ref{n10_MC} Simulation $(N = 2^{10})$ \\
		\ref{n12_MC} Simulation $(N = 2^{12})$
		}
		};

	\end{semilogyaxis}

\end{tikzpicture}%
	\caption{Evaluation of $P_e^{\mathrm{UB}}$ and $P_e^{\mathrm{LB}}$ for polar codes of lengths $N = 256,1024,4096,$ designed for the BEC$(0.5)$ with $\delta = 10^{-6}$.}\label{fig:FERvsR}
\end{figure}

In Figure~\ref{fig:FERvsR}, we present the evaluation of $P_e^{\mathrm{UB}}$ and $P_e^{\mathrm{LB}}$ as a function of $R$ and for $N = 256,1024,2048,$ for a faulty SC decoder with $\delta = 10^{-6}$ and transmission over the BEC$(0.5)$. We also present Monte Carlo simulation results that corroborate our analytical expressions for $P_e^{\mathrm{UB}}$ and $P_e^{\mathrm{LB}}$. We observe that, especially for low rates, $P_e^{\mathrm{UB}}$ and $P_e^{\mathrm{LB}}$ are practically indistinguishable. For rates $R > 0.30$, a difference between the lower bound and the upper bound begins to appear, while for $R > 0.40$ both the upper bound and the lower bound break down and should be replaced by their trivial versions $P_e^{\mathrm{UB}} = 1$ and $P_e^{\mathrm{LB}} = \max _{\mathbf{s} \in \mathcal{A}_n} Z^{(\mathbf{s})}_{n,\delta}$. Moreover, we observe that over a wide range of rates the FER under SC decoding actually increases when the blocklength is increased, contrary to the fault-free case where increasing the blocklength generally decreases the FER. This can be explained if we recall that $Z^{(\mathbf{s})}_{n,\delta} \geq \delta$. Thus, by increasing the blocklength while keeping the rate fixed, we are increasing the number of terms in \eqref{eqn:FER}, and since some of these terms do not decrease beyond some point, the value of the sum can increase.

\subsection{Optimal Blocklength}\label{sec:optblock}
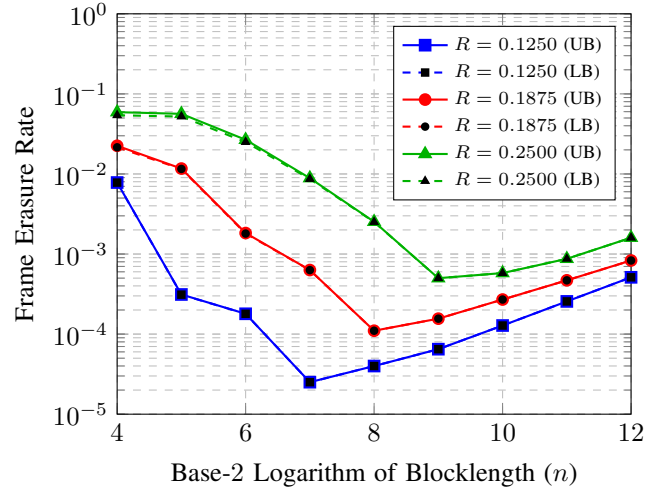
\begin{figure}
	\centering
	\begin{tikzpicture}

	\pgfplotsset{grid style={dashed}}

	\begin{semilogyaxis}[
		width = \figurewidth,
		height = \figureheight,
		xlabel = {Base-$2$ Logarithm of Blocklength ($n$)},
		ylabel = {Frame Erasure Rate},
		xmin = 4, xmax = 12,
		ymin = 1e-5, ymax = 1,
		grid = both,
		legend style={legend pos=north east,font=\scriptsize},
		legend cell align=left,
		legend entries={$R = 0.1250$ (UB), $R = 0.1250$ (LB), $R = 0.1875$ (UB), $R = 0.1875$ (LB), $R = 0.2500$ (UB), $R = 0.2500$ (LB)},
	]		

		% n = 8
		\addplot[blue, thick, solid, mark=square*] table[x index=0, y index = 1] {fig/data/UB_LB_R0.1250_1e-06.dat};
		\addplot[blue, thick, dashed, mark=square*, mark options={solid,black,scale=0.6}] table[x index=0, y index = 2] {fig/data/UB_LB_R0.1250_1e-06.dat};
		
		% n = 10		
		\addplot[red, thick, solid, mark=*] table[x index=0, y index = 1] {fig/data/UB_LB_R0.1875_1e-06.dat};
		\addplot[red, thick, dashed, mark=*, mark options={solid,black,scale=0.6}] table[x index=0, y index = 2] {fig/data/UB_LB_R0.1875_1e-06.dat};

		% n = 12		
		\addplot[darkgreen, thick, solid, mark=triangle*, mark options={scale=1.2}] table[x index=0, y index = 1] {fig/data/UB_LB_R0.2500_1e-06.dat};
		\addplot[darkgreen, thick, dashed, mark=triangle*, mark options={solid,black,scale=0.7}] table[x index=0, y index = 2] {fig/data/UB_LB_R0.2500_1e-06.dat};

	\end{semilogyaxis}

\end{tikzpicture}%
	\caption{Evaluation of $P_e^{\mathrm{UB}}$ and $P_e^{\mathrm{LB}}$ for various blocklengths and code rates and for transmission over a BEC with erasure probability $0.5$ under faulty SC decoding with $\delta = 10^{-6}$.}\label{fig:optN_1e-06}
\end{figure}

An example of the evaluation of $P_{e}^{\mathrm{UB}}$ and $P_{e}^{\mathrm{LB}}$ for $N = 2^n,~n = 4,\hdots,12,$ and code rates $R \in \{0.1250,0.1875,0.2500\}$ (where $K = \lceil RN \rceil$) is shown in Figure~\ref{fig:optN_1e-06} under faulty SC decoding with $\delta = 10^{-6}$ over a BEC$(0.5)$. We observe that the bounds are tight enough in this case so that there always exists a unique $n \in \mathcal{N}$ that satisfies \eqref{eqn:singlen}. Thus, for $R = 0.1250$ the optimal blocklength is $N = 128$, for $R = 0.1875$ the optimal blocklength is $N = 256$, and finally for $R = 0.2500$ the optimal blocklength is $N = 512$. 

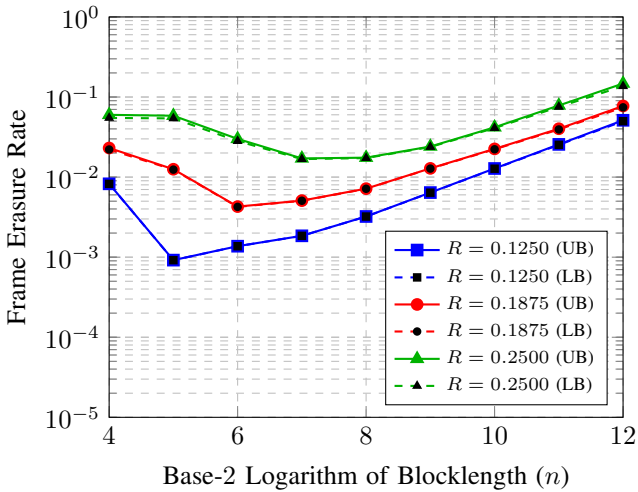
\begin{figure}
	\centering
	\begin{tikzpicture}

	\pgfplotsset{grid style={dashed}}

	\begin{semilogyaxis}[
		width = \figurewidth,
		height = \figureheight,
		xlabel = {Base-$2$ Logarithm of Blocklength ($n$)},
		ylabel = {Frame Erasure Rate},
		xmin = 4, xmax = 12,
		ymin = 1e-5, ymax = 1,
		grid = both,
		legend style={legend pos=south east,font=\scriptsize},
		legend cell align=left,
		legend entries={$R = 0.1250$ (UB), $R = 0.1250$ (LB), $R = 0.1875$ (UB), $R = 0.1875$ (LB), $R = 0.2500$ (UB), $R = 0.2500$ (LB)},
	]		

		% n = 8
		\addplot[blue, thick, solid, mark=square*] table[x index=0, y index = 1] {fig/data/UB_LB_R0.1250_1e-04.dat};
		\addplot[blue, thick, dashed, mark=square*, mark options={solid,black,scale=0.6}] table[x index=0, y index = 2] {fig/data/UB_LB_R0.1250_1e-04.dat};
		
		% n = 10
		\addplot[red, thick, solid, mark=*] table[x index=0, y index = 1] {fig/data/UB_LB_R0.1875_1e-04.dat};
		\addplot[red, thick, dashed, mark=*, mark options={solid,black,scale=0.6}] table[x index=0, y index = 2] {fig/data/UB_LB_R0.1875_1e-04.dat};
		
		% n = 12
		\addplot[darkgreen, thick, solid, mark=triangle*, mark options={scale=1.2}] table[x index=0, y index = 1] {fig/data/UB_LB_R0.2500_1e-04.dat};
		\addplot[darkgreen, thick, dashed, mark=triangle*, mark options={solid,black,scale=0.7}] table[x index=0, y index = 2] {fig/data/UB_LB_R0.2500_1e-04.dat};

	\end{semilogyaxis}

\end{tikzpicture}%
	\caption{Evaluation of $P_e^{\mathrm{UB}}$ and $P_e^{\mathrm{LB}}$ for various blocklengths and code rates and for transmission over a BEC with erasure probability $0.5$ under faulty SC decoding with $\delta = 10^{-4}$.}\label{fig:optN_1e-04}
\end{figure}

Moreover, we present results for $\delta = 10^{-4}$ in Fig.~\ref{fig:optN_1e-04}. We observe that the upper and lower bounds are also tight in this case, but the optimal blocklength for is smaller than for the case of $\delta = 10^{-6}$ for all considered code rates. This is not unexpected, since for a higher $\delta$ the internal decoder erasures will start dominating the error rate at a shorter blocklength.

Thus, we observe that, as the code rate increases, the optimal blocklength generally increases, while as the internal erasure probability $\delta$ increases, the optimal blocklength generally decreases.

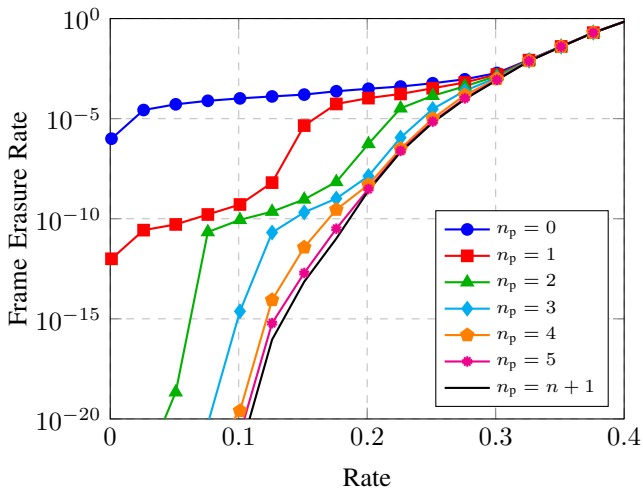
\begin{figure}
	\centering
	\begin{tikzpicture}

	\pgfplotsset{grid style={dashed}}

	\begin{semilogyaxis}[
		cycle list name=exotic,
		width = \figurewidth,
		height = \figureheight,
		xlabel = {Rate},
		ylabel = {Frame Erasure Rate},
		xmin = 0, xmax = 0.4,
		ymin = 1e-20, ymax = 1,
		grid = both,
		legend style={legend pos=south east,font=\scriptsize},
		legend cell align=left,
		legend entries={$n_\text{p}=0$, $n_\text{p}=1$, $n_\text{p}=2$, $n_\text{p}=3$, $n_\text{p}=4$, $n_\text{p}=5$, $n_\text{p} = n+1$},
		cycle list name=color list,
	]		

		% n = 10
		\addplot[blue, thick, solid, mark=*] table[x index=0, y index = 1] {fig/data/FER_n10_stageProt0.dat};
		\addplot[red, thick, solid, mark=square*] table[x index=0, y index = 1] {fig/data/FER_n10_stageProt1.dat};
		\addplot[darkgreen, thick, solid, mark=triangle*, mark options={scale=1.2}] table[x index=0, y index = 1] {fig/data/FER_n10_stageProt2.dat};
		\addplot[cyan, thick, solid, mark=diamond*, mark options={scale=1.2}] table[x index=0, y index = 1] {fig/data/FER_n10_stageProt3.dat};
		\addplot[orange, thick, solid, mark=pentagon*, mark options={scale=1.2}] table[x index=0, y index = 1] {fig/data/FER_n10_stageProt4.dat};
		\addplot[magenta, thick, solid, mark=10-pointed star] table[x index=0, y index = 1] {fig/data/FER_n10_stageProt5.dat};		
		\addplot[black, thick, solid] table[x index=0, y index = 1] {fig/data/FER_n10_stageProtn.dat};		

	\end{semilogyaxis}

\end{tikzpicture}%
	\caption{FER for a polar code of length $N = 1024$ designed for the BEC$(0.5)$ under faulty SC decoding with $\delta = 10^{-6}$ and various numbers of protected decoding levels.}\label{fig:FERvsStageProtection}
\end{figure}

\subsection{Unequal Error Protection}
\begin{figure}[t]
	\centering
	\begin{tikzpicture}

	\pgfplotsset{grid style={dashed}}

	\begin{semilogyaxis}[
		width = \figurewidth,
		height = \figureheight,
		xlabel = {Rate},
		ylabel = {Frame Erasure Rate},
		xmin = 0.05, xmax = 0.5,
		ymin = 1e-20, ymax = 1,
		grid = both,
		legend style={legend pos=south east,font=\scriptsize},
		legend cell align=left,
		legend entries={$N= 2^8$ (faulty), $N= 2^8$ (non-faulty), $N= 2^{10}$ (faulty), $N= 2^{10}$ (non-faulty), $N= 2^{12}$ (faulty), $N= 2^{12}$ (non-faulty)},
	]		

		% n = 8
		\addplot[blue, thick, solid, mark=square*] table {fig/data/UB_n8_protn3.dat};
		\addplot[blue, thick, dashed, mark=square*, mark options={solid,black,scale=0.6}] table {fig/data/UB_n8_noProt.dat};

		% n = 10
		\addplot[red, thick, solid, mark=*] table {fig/data/UB_n10_protn5.dat};
		\addplot[red, thick, dashed, mark=*, mark options={solid,black,scale=0.6}] table {fig/data/UB_n10_noProt.dat};

		% n = 12
		\addplot[darkgreen, thick, solid, mark=triangle*, mark options={scale=1.2}] table {fig/data/UB_n12_protn7.dat};
		\addplot[darkgreen, thick, dashed, mark=triangle*, mark options={solid,black,scale=0.7}] table {fig/data/UB_n12_noProt.dat};

	\end{semilogyaxis}

\end{tikzpicture}%
	\caption{FER for polar codes of length $N = 512, 1024, 2048,$ designed for the BEC$(0.5)$ under faulty SC decoding with $\delta = 10^{-6}$ and $n_{\text{p}} = n-5$ protected decoding levels.}\label{fig:FERvsStageProtectionN}
\end{figure}
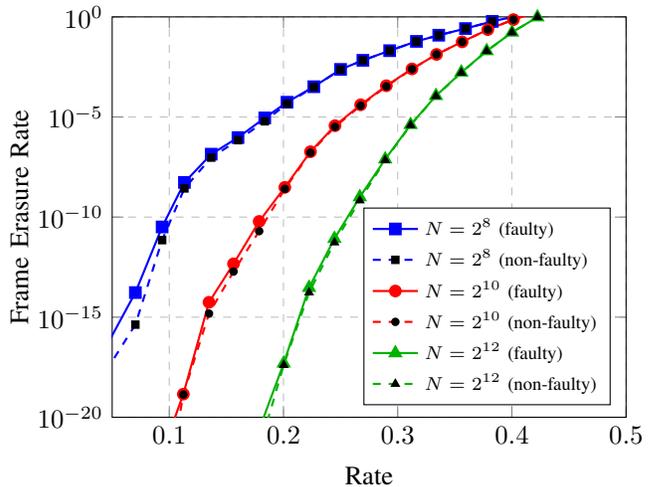

The effect of the partial protection for a finite length code is illustrated in Figure~\ref{fig:FERvsStageProtection}, where we present $P_{e}^{\mathrm{UB}}(\mathcal{A}_n)$ for $N = 2^{10} = 1024$ and $\delta = 10^{-6}$ when $n_{\text{p}} = 0,\hdots,5,$ levels of the tree are protected. To improve readability, we intentionally omit $P_{e}^{\mathrm{LB}}(\mathcal{A}_n)$ from the figure. However, we have already seen that the bounds are tight, especially for low rates, so using only the upper bound is sufficient to illustrate the effect of unequal error protection. We observe that protecting only the root node already improves the performance significantly, especially for the lower rates. When $n_{\text{p}} = 5$, the performance of the faulty SC decoder is almost identical to the non-faulty decoder in the examined FER region and it is remarkable that this performance improvement is achieved by protecting only $\frac{N_{\text{p}}}{N_{\text{PE}}} = \frac{31}{2047} \approx 1.5\%$ of the decoder. Moreover, in Figure~\ref{fig:FERvsStageProtectionN}, we present $P_{e}^{\mathrm{UB}}(\mathcal{A}_n)$ for $N = 512, 1024, 2048,$ and $\delta = 10^{-6}$ with $n_{\text{p}} = n - 5$, so that the protected part for each $N$ is fixed to approximately $1.5\%$ of the decoder. We observe that, contrary to the results of Section~\ref{sec:numerical}, increasing the blocklength actually decreases $P_e(\mathcal{A}_n)$ in the examined FER region, as in the case of the non-faulty decoder.

\section{Conclusion}\label{sec:conclusion}
In this paper, we studied faulty SC decoding of polar codes for the BEC, where the hardware-induced errors are modeled as additional erasures within the decoder. We showed that, under this model, fully reliable communication is not possible at any rate. Furthermore, we showed that, in order for partial ordering of the synthetic channels with respect to the BEC parameter $p$ to hold, the internal erasure probability of the decoder has to be approximately smaller than the erasure probability of the BEC. Moreover, we derived a lower bound on the frame erasure rate and we used this lower bound in order to optimize the blocklength of polar codes under faulty SC decoding. Finally, we proposed an error protection scheme which re-enables asymptotically error-free transmission by protecting only a constant fraction of the decoder. This protection can be implemented using some conventional circuit error-protection mechanism, such as redundancy or increased transistor sizing. Finally, our unequal error protection scheme was shown to significantly improve the performance of the faulty SC decoder for finite-length codes by protecting as little as $1.5\%$ of the decoder.

\ifCLASSOPTIONcaptionsoff
  \newpage
\fi

\ifproofsend
\appendix[Proofs of Properties \ref{property:geqdelta}--\ref{property:covariance}]

\begin{IEEEproof}[Proof of Property~\ref{property:geqdelta}]
For $T^{+}_{\delta}(\epsilon)$, we have
\begin{align}
	\epsilon ^2 + (1-\epsilon ^2)\delta & \geq \delta \Leftrightarrow \\
	(1-\delta)\epsilon ^2 & \geq 0,
\end{align}
which indeed holds for any $\epsilon, \delta \in \left[0,1\right]$. Similarly, for $T^{-}_{\delta}(\epsilon)$, we have
\begin{align}
	2\epsilon - \epsilon ^2 + (1-2\epsilon + \epsilon^2)\delta & \geq \delta \Leftrightarrow \\
	(1-\delta)(2\epsilon - \epsilon ^2) & \geq 0,
\end{align}
which indeed holds for any $\epsilon, \delta \in \left[0,1\right]$.
\end{IEEEproof}

\begin{IEEEproof}[Proof of Property~\ref{property:fixed}]
The above property can easily be shown by solving $T^{+}_{\delta}(\epsilon) = \epsilon$ and $T^{-}_{\delta}(\epsilon) = \epsilon$ for $\epsilon$, respectively, and noting that one solution of $T^{-}_{\delta}(\epsilon) = \epsilon$ is negative.
\end{IEEEproof}

\begin{IEEEproof}[Proof of Property~\ref{prop:nopol}]
This is a direct consequence of Property~\ref{property:geqdelta}, since all $\epsilon _{s,\delta}(q)$ are produced by repeated applications of $T^{+}_{\delta}$ and $T^{-}_{\delta}$ to $\epsilon _{0,\delta} = p$, so that $\epsilon _{s,\delta}(q) \geq \delta,~\forall q \in \mathcal{Q}$.
\end{IEEEproof}

\begin{IEEEproof}[Proof of Property~\ref{property:Tminusplusincreasing}]
(i) $T^{+}_{\delta}(\epsilon)$ can be re-written as
\begin{align}
	T^{+}_{\delta}(\epsilon) 	& = \epsilon ^2 + (1-\epsilon ^2)\delta \\
														& = \epsilon ^2(1-\delta) + \delta.
\end{align}
Thus, for any fixed $\delta \in [0,1]$, $T^{+}_{\delta}(\epsilon)$ is increasing in $\epsilon$ for any $\epsilon \in [0,1]$. Similarly, $T^{-}_{\delta}(\epsilon)$ can be re-written as
\begin{align}
	T^{-}_{\delta}(\epsilon) 	& = 2\epsilon - \epsilon ^2 + (1 - 2\epsilon + \epsilon ^2)\delta \\
														& = (2\epsilon - \epsilon ^2)(1-\delta) + \delta,
\end{align}
which is also increasing in $\epsilon$ for any $\epsilon \in [0,1]$.\\
(ii) Both $T^{-}_{\delta}(\epsilon)$ and $T^{+}_{\delta}(\epsilon)$ are linear functions of $\delta$ with a non-negative slope, so they are increasing $\forall \delta \in \mathbb{R}$.
\end{IEEEproof}

\begin{IEEEproof}[Proof of Property~\ref{prop:monotonicityp}]
The erasure probability of any synthetic channel $\W{s}{s}$ can be calculated by repeated applications of $T^{-}_{\delta}$ and $T^{+}_{\delta}$ starting from $p$ as
\begin{equation}
	\Z{s}{s}(p) = T^{s_s}_{\delta}\left(T^{s_{s-1}}_{\delta}\left(\cdots \left(T^{s_1}_{\delta}(p)\right) \right)\right),
\end{equation}
where $\mathbf{s} = [s_s,s_{s-1},\hdots,s_1]$ and $s_i \in \{+,-\}, i =1,\hdots,s$. Since from Property~\ref{property:Tminusplusincreasing}(i) we know that both $T^{-}_{\delta}(\epsilon)$ and $T^{+}_{\delta}(\epsilon)$ are increasing with respect to $\epsilon$, any composition of the two functions will also be increasing. Thus
\begin{equation}
	\Z{s}{s}(p_2) \leq \Z{s}{s}(p_1) \leq \eta.
\end{equation}
\end{IEEEproof}

\begin{IEEEproof}[Proof of Property~\ref{prop:monotonicityepsilon}]
Similarly to the proof of Property~\ref{prop:monotonicityp}, the proof stems directly from the monotonicity of $T^{-}_{\delta}(\epsilon)$ and $T^{+}_{\delta}(\epsilon)$ with respect to $\delta$ shown in Property~\ref{property:Tminusplusincreasing}(ii).
\end{IEEEproof}

\begin{IEEEproof}[Proof of Property~\ref{property:covariance}]
For simpler notation, let us define $X' \triangleq X + (1-X)\Delta_1$ and $Y' \triangleq Y + (1-Y)\Delta_2$. We then have
\begin{align}
	\mathrm{cov}\left[X',Y'\right]	& = \mathbb{E}[X'Y'] - \mathbb{E}[X']\mathbb{E}[Y'] \\
									& = \mathbb{E}[(1-\Delta_1)X + \Delta_1)((1-\Delta_2)Y + \Delta_2)] \nonumber \\
									& - \mathbb{E}[(1-\Delta_1)X + \Delta_1]\mathbb{E}[(1-\Delta_2)Y + \Delta_2] \\
									& \stackrel{(*)}{=} \Ex{(1-\Delta_1)(1-\Delta_2)}(\mathbb{E}[XY] - \mathbb{E}[X]\mathbb{E}[Y]) \\
									& \stackrel{(**)}{=} (1-\delta)^2\mathrm{cov}\left[X,Y\right],
\end{align}
where for $(*)$ we have used the independence of $\Delta_1$ and $\Delta_2$ from $X$ and $Y$, while for $(**)$ we have used the independence between $\Delta_1$ and $\Delta_2$.
\end{IEEEproof}

\fi



%\bibliographystyle{IEEEtran}
% http://www.ctan.org/tex-archive/macros/latex/contrib/supported/IEEEtran/bibtex
\bibliographystyle{IEEEtran}
\bibliography{IEEEabrv,bibliography}


%\end{IEEEbiography}


%\newpage



%\vfill


\end{document}